\documentclass[11pt]{article}

\usepackage[final]{latex/acl}

\usepackage{times}
\usepackage{latexsym}
\usepackage[T1]{fontenc}
\usepackage[utf8]{inputenc}
\usepackage{microtype}
\usepackage{inconsolata}
\usepackage{graphicx}

\title{LintEval: Benchmarking LLMs on Linter-Guided File Refactoring}

\author{
  Tongwei Deng\textsuperscript{1,2},
  Jinyu Zhu\textsuperscript{1,2},
  Yue Cao\textsuperscript{3},
  Chenxu Wang\textsuperscript{1,2},
  Yanzhen Zou\textsuperscript{1,2},
  \setcounter{footnote}{1}
  Bing Xie\thanks{Corresponding author.}\textsuperscript{1,2} \\
  \textsuperscript{1}School of Computer Science, Peking University \\
  \textsuperscript{2}Key Lab of HCST (PKU), MOE; SCS, Peking University \\
  \textsuperscript{3}School of Software Engineering, Beijing Jiaotong University \\
  \texttt{\{dtw,2601112166,2400013225\}@stu.pku.edu.cn},
  \texttt{23301030@bjtu.edu.cn}, \\
  \texttt{\{zouyz,xiebing\}@pku.edu.cn}
}

\usepackage[most]{tcolorbox}
\usepackage{pifont}
\definecolor{lightgray}{gray}{0.95}
\definecolor{darkgray}{gray}{0.2}
\newcommand{\cmark}{\textcolor{green}{\ding{51}}}
\newcommand{\xmark}{\textcolor{red}{\ding{55}}}
\usepackage{amssymb}
\usepackage{amsmath}
\usepackage{multirow}
\usepackage{makecell}
\usepackage{booktabs}
\usepackage{adjustbox}
\usepackage[table]{xcolor}

\newtcolorbox{summarybox}{
    enhanced, % 必须开启 enhanced 才能使用 borderline
    colback=lightgray,
    colframe=lightgray,
    boxrule=0pt,
    left=1em,
    right=1em,
    top=0.5em,
    bottom=0.5em,
    borderline west={3pt}{0pt}{darkgray}, % 更深的颜色提升可视性
    sharp corners,
    breakable
}

\usepackage[most]{tcolorbox}
\usepackage{listings}
\usepackage{textcomp}

\lstdefinestyle{promptstyle}{
  basicstyle=\ttfamily\scriptsize,
  columns=fullflexible,
  breaklines=true,
  breakatwhitespace=false,
  keepspaces=true,
  showstringspaces=false,
  upquote=true
}

\newtcblisting{promptbox}[1][]{
  enhanced,
  width=\columnwidth,
  colback=gray!3,
  colframe=black!35,
  boxrule=0.4pt,
  arc=1pt,
  left=3pt,
  right=3pt,
  top=3pt,
  bottom=3pt,
  fonttitle=\footnotesize\bfseries,
  listing only,
  listing options={style=promptstyle},
  #1
}

\begin{document}

\maketitle

\begin{abstract}

Linters are widely used in code maintenance and real-time IDE editing to surface lint alerts related to performance, style, error-prone, and other patterns in source files. 
However, existing code benchmarks do not explicitly evaluate linter-guided file refactoring, where models must reduce lint alerts in source files with correct initial functionality while satisfying functionality preservation. 
We introduce LintEval, a multilingual benchmark for linter-guided file refactoring across C++, Python, and Java, comprising 13,061 programs with correct initial functionality and 171,662 lint alerts extracted using language-specific linters. 
LintEval jointly evaluates functionality preservation, lint alert reduction, new alert avoidance, and edit minimality. 
Experiments reveal a persistent trade-off: closed-source commercial models reach up to 90.4\% Reduction@P but achieve only 67.1\%--68.3\% Pass@1, while several open-source small models show low or negative Reduction@P. 
A controlled fine-tuning study further shows that supervision format matters: refactoring-oriented SFT raises Qwen2.5-Coder-7B Reduction@P from 24.7\% to 76.0\%. 
LintEval enables controlled evaluation of models that improve source files under functionality preservation.

\end{abstract}

\noindent\textbf{Code:} \url{https://github.com/6643789wsx/LintEval}

\section{Introduction}

Large Language Models (LLMs) have advanced many software-engineering tasks \citep{yang2025code_intelligence_survey, hui2024qwen2, wang2024xcode}. Existing code benchmarks assess them through code generation from specifications \citep{chen2021evaluating, ge2024survey, tong-zhang-2024-codejudge, huang2024da-code, uniyal2024onetomany, xinyi2024conline, huang2022execution}, automated debugging and repair \citep{zhang2024survey, li2024debug, tian2024debugbench, bui2022detect}, and non-functional requirements or refactoring beyond functional correctness \citep{manav2024nofuneval, gautam2025refactorbench}.

Within the scope of functionality-preserving refactoring, linter-guided file refactoring remains under-explored despite its direct connection to developer workflows.
Linters identify performance inefficiencies, style violations, and fine-grained risks within source files, and resolving these file-level alerts helps maintain code health and reduce technical debt \citep{christakis2016what, tomasdottir2020adoption, zampetti2017how, beller2016analyzing, hansson2024ghostechoes}.
Because many alerts require context-sensitive changes, they provide a controlled setting for testing whether LLMs can resolve code-quality issues beyond fixed rule-based transformations \citep{venkatesh2024emergence, AlBatainehM22}.
However, despite the increasing use of static-analysis and code-quality feedback to improve model outputs \citep{greta2024helping, RL_CQ_GRPO}, existing benchmarks fail to jointly evaluate functionality preservation, lint alert reduction, new alert avoidance, and edit minimality.

% Linter-guided file refactoring presents four key challenges: (1) functionality preservation: lint alerts must be resolved without introducing regressions or compromising program semantics; (2) broad spectrum of alerts: lint alerts range from simple style issues to complex semantic pitfalls such as resource leaks; (3) new alert avoidance: refactoring must not introduce collateral alerts that negate the benefits of the original resolution; and (4) minimal modifications: edits should be localized to specific segments rather than performing unnecessary global rewrites.

\begin{figure*}[t]
     \centering
     \includegraphics[width=\linewidth]{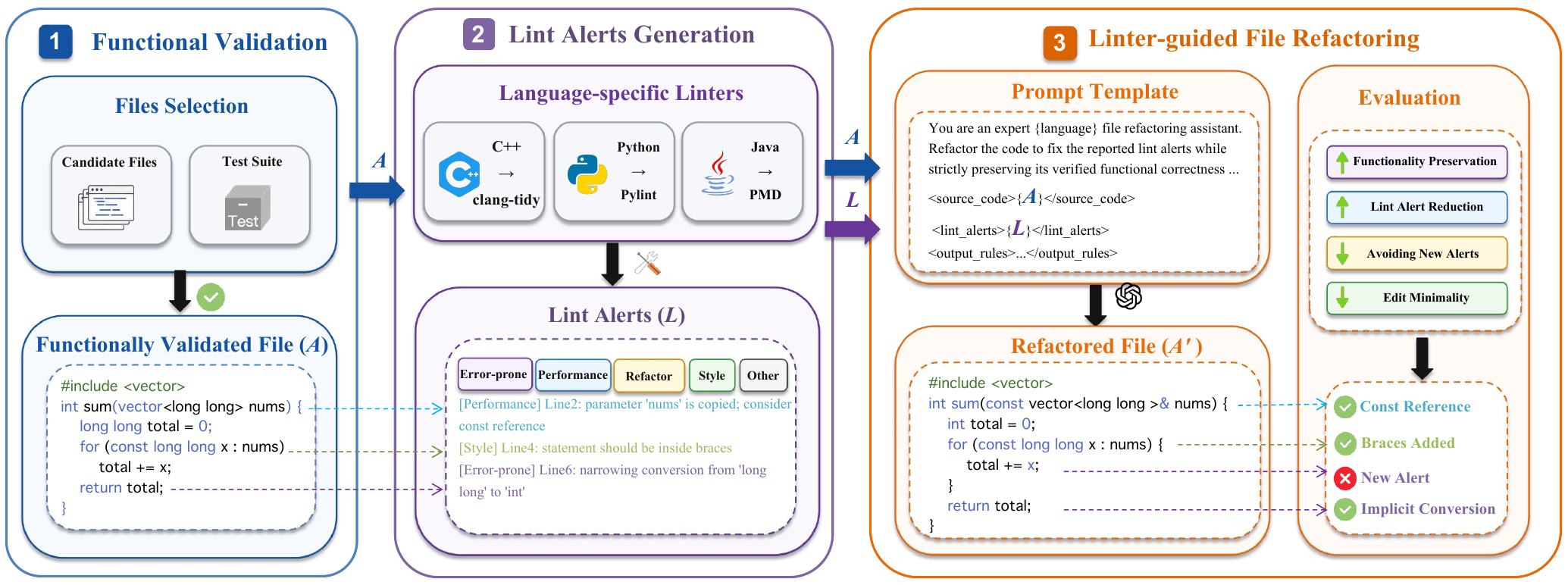}
    \caption{\textbf{The LintEval framework for linter-guided file refactoring.} The pipeline systematically extracts lint alerts ($L$) from functionally correct programs ($A$) across multiple programming languages to evaluate LLM-driven revision. The resulting refactored file ($A'$) is evaluated across four dimensions, where quality metrics are strictly conditionalized on functionality preservation to ensure new alert avoidance and edit minimality.}
    \label{fig:pipeline}
 \end{figure*}

We introduce \textbf{LintEval}, a large-scale, multilingual benchmark for linter-guided file refactoring (Figure~\ref{fig:pipeline}, Table~\ref{tab:benchmark_comparison}). To support rigorous functionality preservation, we curate 13,061 files across C++, Python, and Java from CodeContests+ \citep{codecontestsplus2025}, using algorithmic tasks not to mimic all production settings but to obtain high-coverage test oracles that detect subtle functionality preservation failures. Within these files, we identify 171,662 lint alerts.

We evaluate representative closed-source commercial and open-source models alongside rule-based baselines, and find that linter-guided file refactoring remains challenging for current models. Rule-based baselines resolve only 25.9\% of lint alerts, while closed-source commercial models prioritize lint alert reduction at the cost of functionality preservation, introducing functional regressions in over 31\% of samples; a pronounced cross-lingual disparity also persists (e.g., DeepSeek-v3.2 falls from 76.6\% Reduction@P on C++ to 63.7\% on Java). A controlled fine-tuning study shows that explicit lint alerts in the training format, rather than exposure to cleaner code alone, are central to this capability: refactoring-oriented supervision raises Qwen2.5-Coder-7B from 24.7\% to 76.0\% Reduction@P.

\begin{table}[t]
\centering
\small
\begin{adjustbox}{max width=\columnwidth}
\begin{tabular}{l c c c}
\toprule
\textbf{Benchmark} & \textbf{Initial Functionality} & \textbf{Multilingual} & \textbf{Lint Alerts} \\
\midrule
\multicolumn{4}{l}{\textit{Code Generation}} \\
HumanEval    & None & \xmark & \xmark \\
MBPP         & None & \xmark & \xmark \\
CodeContests & None & \cmark & \xmark \\
\midrule
\multicolumn{4}{l}{\textit{Bug Repair}} \\
Defects4J    & Incorrect    & \xmark & \xmark \\
QuixBugs     & Incorrect    & \xmark & \xmark \\
DebugBench   & Incorrect    & \xmark & \xmark \\
\midrule
\multicolumn{4}{l}{\textit{Code Refactoring}} \\
RefactorBench & Correct     & \xmark & \xmark \\
\rowcolor[gray]{0.9} \textbf{LintEval (Ours)} & \textbf{Correct} & \textbf{\cmark} & \textbf{\cmark} \\
\bottomrule
\end{tabular}
\end{adjustbox}
\caption{Comparison with existing benchmarks.}
\label{tab:benchmark_comparison}
\end{table}

Core contributions are summarized as follows:
\begin{itemize}
    \item \textbf{Benchmark:} We introduce \textbf{LintEval}, a large-scale, multilingual benchmark for linter-guided file refactoring, featuring 13,061 files and 171,662 lint alerts to evaluate lint alert reduction by LLMs under strict functionality preservation.
    \item \textbf{Empirical Analysis:} Through extensive evaluation of representative models and training settings, we map model boundaries: closed-source commercial models prioritize lint alert reduction over functionality preservation, while open-source models show cross-language and alert-category weaknesses.
    \item \textbf{Fine-Tuning Study:} We demonstrate that while generation-oriented training does not reliably improve linter-guided file refactoring, refactoring-oriented data formatting substantially narrows the gap between open-source models and closed-source commercial models on this task.
\end{itemize}

\section{LintEval Construction}
\label{sec:construction}
 
LintEval is built through a controlled pipeline for evaluating linter-guided file refactoring. As illustrated in Figure~\ref{fig:pipeline}, we curate initial source files exclusively from programs with correct initial functionality under high-coverage test oracles, establishing a reliable baseline to validate subsequent functionality preservation. We then deploy industry-standard static analyzers across C++, Python, and Java, normalizing heterogeneous lint alerts into a unified taxonomy.

\subsection{Data Sourcing and Functional Validation}
\label{sec:data_sourcing}

We source files and test suites from \textit{CodeContests+} \citep{codecontestsplus2025}, which expands test volume to uncover subtle functional regressions.
Each problem provides 50--200 test cases, giving high-coverage oracles that establish correct initial functionality and detect functionality preservation failures after refactoring.

\subsection{Multi-Language Alert Extraction}
\label{sec:alert_extraction}

LintEval identifies lint alerts by running version-pinned linters directly on files that pass the provided test suites.
This design avoids model-generated lint alerts and makes the extraction protocol reproducible.
Specifically, we deploy \texttt{clang-tidy} (22.1.0) for C++, \texttt{Pylint} (4.0.5) for Python, and \texttt{PMD} (7.12.0) for Java.
The comprehensive sets of active rules and detailed configurations are provided in Appendix~\ref{sec:appendix_execution_env} (Table~\ref{app:execution_env}) to support reproducibility.

\subsection{Benchmark Instance Construction}
Each benchmark instance assembles the source code and its extracted lint alerts into a standardized prompt (Figure~\ref{fig:pipeline}) that instructs the model to reduce the provided lint alerts while satisfying functionality preservation. The full prompt is provided in Appendix~\ref{sec:appendix_prompt}.

% \begin{figure}[t]
% \centering
% \begin{promptbox}[title={Prompt Template Layout}]
% You are an expert {language} file refactoring assistant. 

% Refactor the existing file to fix the reported lint alerts while strictly preserving its verified functional correctness.

% <requirements>
% {requirements}
% </requirements>

% <source_code>
% {A}
% </source_code>

% <lint_alerts>
% {L}
% </lint_alerts>

% {output_rules}
% \end{promptbox}
% \caption{Prompt template schema for linter-guided file refactoring, mapping input source programs ($A$) and corresponding lint alerts ($L$) into structured evaluation contexts.}
% \label{fig:prompt-template}
% \end{figure}

\subsection{Dataset Characteristics and Statistics}
\label{sec:dataset_stats}

\begin{figure}[t]
    \centering
    \includegraphics[width=\columnwidth]{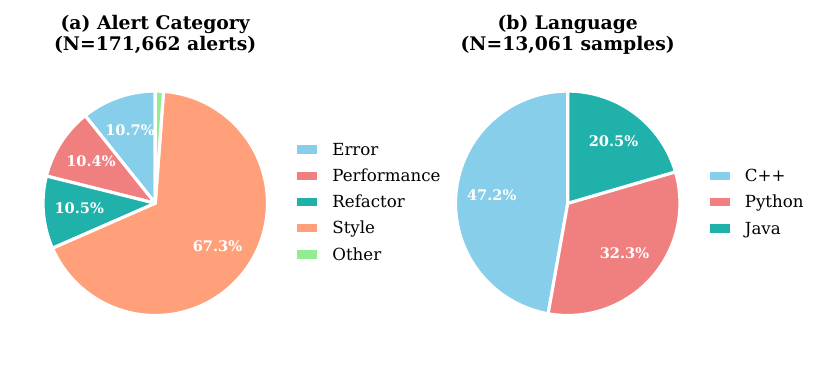}
    \caption{Distribution of lint alert categories ($N{=}171{,}662$ alerts) and programming languages ($N{=}13{,}061$ samples) in LintEval.}
    \label{fig:dataset_distribution}
\end{figure}

As summarized in Table~\ref{tab:dataset_statistics}, LintEval comprises 13,061 samples spanning 3,999 unique problems across C++ (6,165), Python (4,221), and Java (2,675).
Average program length is 61.0 LOC with a density of 13.1 alerts per sample.
As illustrated in Figure~\ref{fig:dataset_distribution}, the lint alert categories are dominated by Style (67.3\%), while Error-prone, Performance, and Refactor domains collectively account for 31.6\%.

% ============================================================
% Placement: Main Text (Section 2 - Dataset)
% ============================================================
% T1: Dataset Statistics
\begin{table}[t]
\centering
\small
\begin{tabular}{lr}
\toprule
\textbf{Metric} & \textbf{Value} \\
\midrule
Unique problems & 3,999 \\
Samples & 13,061 \\
Total lint alerts & 171,662 \\
Avg. alerts per sample & 13.1 \\
Avg. LOC per sample & 61.0 \\
Unique alert types & 396 \\
\midrule
\multicolumn{2}{l}{\textit{Language counts}} \\
C++ & 6,165 \\
Python & 4,221 \\
Java & 2,675 \\
\midrule
\multicolumn{2}{l}{\textit{Alert category distribution}} \\
Error-prone & 18,434 \\
Performance & 17,776 \\
Refactor & 17,977 \\
Style & 115,458 \\
Other & 2,017 \\
\bottomrule
\end{tabular}
\caption{Dataset statistics for \textbf{LintEval}. The dataset contains 13,061 competitive-programming samples across three languages, each annotated with lint alerts.}
\label{tab:dataset_statistics}
\end{table}

\subsection{Taxonomy Normalization}
\label{sec:taxonomy}

Given the disparate output formats of different linters, we normalize all detected alerts into a unified taxonomy to facilitate cross-language evaluation. We categorize lint alerts into five dimensions: Error-prone, Performance, Refactor, Style, and Other. Table~\ref{tab:dataset_statistics} presents the alert counts per category. The detailed cross-language mapping is provided in Appendix~\ref{sec:appendix_taxonomy}.

For evaluation, two alerts are considered identical if they share the same rule identifier and sanitized text content. We explicitly exclude line numbers from this matching process to prevent interference from line shifts caused by code edits.

\section{Problem Formulation}
\label{sec:formulation}

\subsection{Task Definition}
Linter-guided file refactoring targets a source file $A$ with correct initial functionality, paired with a set of lint alerts $L$ detected on it, and a test suite $T$ that $A$ already passes.
Given the tuple $(A, L, T)$, the model produces a refactored file $A' = f(A, L)$ subject to four evaluation dimensions: functionality preservation, lint alert reduction, new alert avoidance, and edit minimality.
Unlike code generation or automated program repair, this task enforces functionality preservation as a strict prerequisite rather than an objective to be recovered.
A benchmark instance is formatted as a tuple $(A_i, L_i, T_i) \in \mathcal{D}$, where $\mathcal{D} = \{(A_i, L_i, T_i)\}_{i=1}^{N}$.

\subsection{Evaluation Metrics}
\label{sec:metrics}
Let $L'_i$ denote the residual lint alerts within the refactored file $A'_i$.
We define the functionally correct subset $S_{\text{pass}}$ via dynamic verification as:
\begin{equation}
S_{\text{pass}} = \{\, i \mid \text{Pass}(A'_i, T_i) = 1 \,\}.
\end{equation}
For each individual instance, we define the following foundational indicators: (1) $\text{Fixed}_i = |L_i \setminus L'_i|$, representing the volume of resolved original alerts (counted by rule identifier and sanitized text after excluding line numbers), (2) $\text{New}_i = |L'_i \setminus L_i|$, representing the volume of newly introduced alerts (counted similarly), and (3) $\text{Lines}_i$, denoting the absolute sum of added and deleted lines.
Pass@1 is evaluated globally over $\mathcal{D}$, whereas all quality metrics carry the suffix \texttt{@P}, denoting that they are computed only over $S_{\text{pass}}$; this excludes alert reduction achieved at the cost of functional regression.

\paragraph{Functionality Preservation (Pass@1).}
The fraction of refactored files that pass the same test suite as the initial file:
\begin{equation}
\text{Pass@1} = \frac{|S_{\text{pass}}|}{N}.
\end{equation}

\paragraph{Lint Alert Reduction (Reduction@P).}
Aggregating at the corpus level prevents files with few initial alerts from biasing the mean; Reduction@P is the net fraction of original alerts eliminated over the functionally correct subset:
\begin{equation}
\text{Reduction@P} = \frac{\sum_{i \in S_{\text{pass}}} \bigl( \text{Fixed}_i - \text{New}_i \bigr)}{\sum_{i \in S_{\text{pass}}} |L_i|}.
\end{equation}
A negative value indicates that the model introduces more new alerts than it resolves.

\begin{table*}[t]
\centering
\small
\begin{tabular}{l ccccc}
\toprule
\textbf{Model} & \textbf{Pass@1} ($\uparrow$) & \textbf{Reduction@P} ($\uparrow$) & \textbf{New@P} ($\downarrow$) & \textbf{$\Delta$Line@P} ($\downarrow$) & \textbf{LinePerFix@P} ($\downarrow$) \\
\midrule
Copy-Original & 100.0\% & 0.0\% & 0.0\% & 0.0 & -- \\
Rule-Based Fixers & \textbf{76.3\%} & 25.9\% & \textbf{0.6\%} & 71.4 & 51.7 \\
\midrule
GPT-5.2 & 67.1\% & 87.8\% & 7.3\% & 71.7 & 16.9 \\
GPT-5.4-mini & 68.3\% & \textbf{90.4\%} & 5.6\% & 60.2 & 14.2 \\
\midrule
DeepSeek-v3.2 & 57.6\% & 73.0\% & 6.7\% & 45.4 & 13.8 \\
Qwen2.5-Coder-32B & 61.7\% & 57.1\% & 8.1\% & 33.3 & 12.6 \\
Qwen2.5-Coder-7B & 53.0\% & 24.7\% & 12.7\% & 32.5 & 12.7 \\
Qwen2.5-Coder-3B & 52.0\% & 9.5\% & 11.2\% & 27.9 & 12.8 \\
Qwen2.5-Coder-0.5B & 48.8\% & $-1.2\%$ & 10.5\% & \textbf{17.2} & \textbf{9.5} \\
Llama-3.2-3B & 38.8\% & 19.8\% & 13.5\% & 29.8 & 12.8 \\
Llama-3.2-1B & 33.7\% & 0.5\% & 13.8\% & 23.5 & 13.2 \\
\bottomrule
\end{tabular}%
\caption{Overall results on \textbf{LintEval}. Reduction@P, New@P, $\Delta$Line@P, and LinePerFix@P are computed only over passed instances. Best results per column are highlighted in \textbf{boldface}.}
\label{tab:overall_performance}
\end{table*}

\paragraph{new alert avoidance (New@P).}
Sharing the corpus-level denominator with Reduction@P, New@P measures the fraction of introduced new alerts relative to the baseline alert volume:
\begin{equation}
\text{New@P} = \frac{\sum_{i \in S_{\text{pass}}} \text{New}_i}{\sum_{i \in S_{\text{pass}}} |L_i|}.
\end{equation}
Because Reduction@P and New@P use the same denominator, Reduction@P decomposes into fixed original alerts minus new alerts:
\begin{equation}
\text{Reduction@P} = \frac{\sum_{i \in S_{\text{pass}}} \text{Fixed}_i}{\sum_{i \in S_{\text{pass}}} |L_i|} - \text{New@P}.
\end{equation}
Lower New@P values indicate stronger performance in new alert avoidance during modification.

\paragraph{Edit Minimality ($\Delta$Line@P).}
The average number of added and deleted lines per functionally correct refactoring instance:
\begin{equation}
\Delta\text{Line@P} = \frac{1}{|S_{\text{pass}}|} \sum_{i \in S_{\text{pass}}} \text{Lines}_i .
\end{equation}

\paragraph{Edit Minimality (LinePerFix@P).}
The amortized edit cost per unit of successfully resolved original alert:
\begin{equation}
\text{LinePerFix@P} = \frac{\sum_{i \in S_{\text{pass}}} \text{Lines}_i}{\sum_{i \in S_{\text{pass}}} \text{Fixed}_i} .
\end{equation}
\label{sec:lineperfix}

\section{Experiments}
\label{sec:experiments}

% ============================================================
% Table: Failure Case Analysis
% ============================================================
\begin{table*}[t]
\centering
\small
\resizebox{\linewidth}{!}{
\begin{tabular}{l cccc cccc}
\toprule
& \multicolumn{4}{c}{\textbf{Pass Cases}} & \multicolumn{4}{c}{\textbf{Failure Cases}} \\
\cmidrule(lr){2-5} \cmidrule(lr){6-9}
\textbf{Model} & \textbf{Reduction@P} & \textbf{New@P} & \textbf{$\Delta$Line@P} & \textbf{LinePerFix@P} & \textbf{Reduction@F} & \textbf{New@F} & \textbf{$\Delta$Line@F} & \textbf{LinePerFix@F} \\
\midrule
GPT-5.2 & 87.8\% & 7.3\% & 71.7 & 16.9 & 86.8\% \textcolor{red!70!black}{$\downarrow$} & 6.3\% \textcolor{green!70!black}{$\downarrow$} & 65.5 \textcolor{green!70!black}{$\downarrow$} & 17.1 \textcolor{red!70!black}{$\uparrow$} \\
GPT-5.4-mini & \textbf{90.4\%} & \textbf{5.6\%} & 60.2 & 14.2 & \textbf{89.2\%} \textcolor{red!70!black}{$\downarrow$} & \textbf{4.6\%} \textcolor{green!70!black}{$\downarrow$} & 69.0 \textcolor{red!70!black}{$\uparrow$} & 13.9 \textcolor{green!70!black}{$\downarrow$} \\
\midrule
DeepSeek-v3.2 & 73.0\% & 6.7\% & 45.4 & 13.8 & 41.6\% \textcolor{red!70!black}{$\downarrow$} & 6.5\% \textcolor{green!70!black}{$\downarrow$} & 23.6 \textcolor{green!70!black}{$\downarrow$} & 12.4 \textcolor{green!70!black}{$\downarrow$} \\
Qwen2.5-Coder-32B & 57.1\% & 8.1\% & 33.3 & 12.6 & 45.3\% \textcolor{red!70!black}{$\downarrow$} & 13.7\% \textcolor{red!70!black}{$\uparrow$} & 30.4 \textcolor{green!70!black}{$\downarrow$} & \textbf{10.4} \textcolor{green!70!black}{$\downarrow$} \\
Qwen2.5-Coder-0.5B & -1.2\% & 10.5\% & \textbf{17.2} & \textbf{9.5} & -169.1\% \textcolor{red!70!black}{$\downarrow$} & 11.7\% \textcolor{red!70!black}{$\uparrow$} & 13.4 \textcolor{green!70!black}{$\downarrow$} & 131.7 \textcolor{red!70!black}{$\uparrow$} \\
Qwen2.5-Coder-3B & 9.5\% & 11.2\% & 27.9 & 12.8 & -155.3\% \textcolor{red!70!black}{$\downarrow$} & 9.2\% \textcolor{green!70!black}{$\downarrow$} & 17.7 \textcolor{green!70!black}{$\downarrow$} & 31.3 \textcolor{red!70!black}{$\uparrow$} \\
Qwen2.5-Coder-7B & 24.7\% & 12.7\% & 32.5 & 12.7 & -88.2\% \textcolor{red!70!black}{$\downarrow$} & 13.6\% \textcolor{red!70!black}{$\uparrow$} & 18.8 \textcolor{green!70!black}{$\downarrow$} & 17.3 \textcolor{red!70!black}{$\uparrow$} \\
Llama-3.2-1B & 0.5\% & 13.8\% & 23.5 & 13.2 & -219.6\% \textcolor{red!70!black}{$\downarrow$} & 14.3\% \textcolor{red!70!black}{$\uparrow$} & \textbf{8.4} \textcolor{green!70!black}{$\downarrow$} & 34.4 \textcolor{red!70!black}{$\uparrow$} \\
Llama-3.2-3B & 19.8\% & 13.5\% & 29.8 & 12.8 & -180.2\% \textcolor{red!70!black}{$\downarrow$} & 23.2\% \textcolor{red!70!black}{$\uparrow$} & 10.5 \textcolor{green!70!black}{$\downarrow$} & 14.7 \textcolor{red!70!black}{$\uparrow$} \\
\bottomrule
\end{tabular}
}
\caption{Editing behavior of passed and failed refactoring attempts. \texttt{@P} metrics are computed over $S_{\text{pass}}$; \texttt{@F} metrics are computed over failed instances for diagnostic analysis only. Arrows indicate whether the failure-case value is higher or lower than the pass-case value. \textbf{Bold} marks the best value per column.}
\label{tab:failure_analysis}
\end{table*}

% ============================================================
% Placement: Main Text (Section 4.2 - Language Analysis)
% ============================================================
% T3: Language-wise Performance (Main Text - Selected Metrics)
% Pass@1 and Reduction@P by programming language
\begin{table*}[t]
\centering
\small
\begin{adjustbox}{max width=\textwidth}
\begin{tabular}{lcccccc}
\toprule
& \multicolumn{2}{c}{\textbf{C++}} & \multicolumn{2}{c}{\textbf{Python}} & \multicolumn{2}{c}{\textbf{Java}} \\
\cmidrule(lr){2-3} \cmidrule(lr){4-5} \cmidrule(lr){6-7}
\textbf{Model} & Pass@1 & Reduction@P & Pass@1 & Reduction@P & Pass@1 & Reduction@P \\
\midrule
GPT-5.2 & 67.2\% & 94.9\% & 64.5\% & 84.4\% & \textbf{76.5\%} & 77.5\% \\
GPT-5.4-mini & \textbf{69.7\%} & \textbf{95.6\%} & \textbf{65.5\%} & \textbf{86.5\%} & 72.5\% & \textbf{81.7\%} \\
\midrule
DeepSeek-v3.2 & 68.0\% & 76.6\% & 46.8\% & 70.6\% & 51.0\% & 63.7\% \\
Qwen2.5-Coder-32B & 64.3\% & 58.6\% & 58.1\% & 63.5\% & 62.7\% & 37.2\% \\
Qwen2.5-Coder-0.5B & 63.9\% & 1.5\% & 32.5\% & -7.2\% & 41.2\% & -3.6\% \\
Qwen2.5-Coder-3B & 66.8\% & 9.5\% & 35.0\% & 10.1\% & 49.0\% & 8.5\% \\
Qwen2.5-Coder-7B & 67.6\% & 27.6\% & 35.5\% & 19.0\% & 52.9\% & 21.6\% \\
Llama-3.2-1B & 44.7\% & 5.2\% & 23.2\% & -23.6\% & 23.5\% & 18.5\% \\
Llama-3.2-3B & 50.8\% & 18.8\% & 23.6\% & 13.2\% & 41.2\% & 31.4\% \\
\bottomrule
\end{tabular}
\end{adjustbox}
\caption{Pass@1 and Reduction@P by programming language on LintEval. Best results per column are in \textbf{bold}.}
\label{tab:language_performance}
\end{table*}

\subsection{Experimental Setup}

\paragraph{Models.}
We evaluate two closed-source commercial models (\texttt{GPT-5.2}, \texttt{GPT-5.4-mini}) and the following open-source models on the \textbf{LintEval} benchmark: \texttt{DeepSeek-v3.2}, \texttt{Qwen2.5-Coder-0.5B/3B/7B/32B}, and \texttt{Llama-3.2-1B/3B}. To broaden the scale and recency coverage, we additionally evaluate \texttt{Qwen3-4B/8B}, \texttt{Qwen3.5-9B}, and \texttt{Qwen3.6-27B/35B-A3B} under the same single-turn, zero-shot protocol without fine-tuning. The appendix also reports SFT robustness checks with \texttt{DeepSeek-Coder-6.7B-Instruct}. Thus, the evaluation includes multiple small open-source models from 0.5B to 4B, alongside larger checkpoints, rather than relying on a single model scale. All models use the fixed greedy decoding configuration reported in Appendix~\ref{sec:appendix_decoding} and an identical prompt to ensure a standardized comparison. SFT variants are analyzed separately in Section~\ref{sec:sft}. We use Qwen2.5-Coder as the controlled SFT backbone to hold the backbone fixed across supervision formats.

\paragraph{Rule-Based Baselines.}
We compare against two control references. \textbf{Copy-Original} returns the source code unchanged, establishing a zero-edit upper bound on functionality preservation. \textbf{Rule-Based Baselines} apply developer-standard tools: \texttt{ruff --fix} (Python), \texttt{clang-tidy --fix} (C++), and \texttt{google-java-format} (Java).\footnote{\texttt{google-java-format} targets only formatting alerts, so the Java rule-based baseline is conservative on non-formatting categories.}

\subsection{Main Results}

\paragraph{Model Recency.}
We additionally evaluate newer Qwen checkpoints under the same single-turn, zero-shot protocol without fine-tuning. This comparison tests whether the main findings depend on the choice of open-source backbone and reports both functionality preservation and lint alert reduction; the results are shown in Table~\ref{tab:new_qwen_check}.

\begin{table}[t]
\centering
\small
\begin{tabular}{lrr}
\toprule
Model & Pass@1 & Reduction@P \\
\midrule
Qwen3-4B & 45.8 & 32.0 \\
Qwen3-8B & 33.3 & 70.7 \\
Qwen3.5-9B & 37.5 & 31.3 \\
Qwen3.6-27B & 64.3 & 62.0 \\
Qwen3.6-35B-A3B & 65.3 & 61.5 \\
\bottomrule
\end{tabular}
\caption{Newer Qwen checkpoints under the same single-turn, zero-shot protocol without fine-tuning. No checkpoint dominates both functionality preservation and lint alert reduction.}
\label{tab:new_qwen_check}
\end{table}

\paragraph{Functionality Preservation and Lint Alert Reduction}
Linter-guided file refactoring exposes three behavioral regimes that no single model category dominates simultaneously (Table~\ref{tab:overall_performance}).
Excluding the no-op Copy-Original control, rule-based baselines achieve the highest functionality preservation ($76.3\%$ Pass@1) but resolve only $25.9\%$ of lint alerts, suggesting that fixed rule coverage leaves many lint alerts unresolved under this protocol.
Additional-analyzer and tool-coverage audits confirm that results depend partly on analyzer--repair-tool coverage; full diagnostics are reported in Appendix~\ref{sec:appendix_camera_ready}.
Closed-source commercial models achieve the highest lint alert reduction (e.g., \texttt{GPT-5.4-mini} at $90.4\%$ Reduction@P), but their Pass@1 remains $67.1\%$--$68.3\%$, indicating a clear trade-off with functionality preservation.
Several open-source models show low or negative lint alert reduction (as low as $-1.2\%$ for \texttt{Qwen2.5-Coder-0.5B}), especially when edits introduce more new alerts than they resolve.
Across rule-based baselines, closed-source commercial models, and open-source models, no method jointly optimizes all four dimensions; these results suggest that the trade-off is not solely an artifact of any single method, but a recurring constraint of linter-guided file refactoring.

\paragraph{Edit Minimality}
Although rule-based baselines make the largest edits ($\Delta$Line@P $= 71.4$) for only moderate alert reduction, closed-source commercial models achieve substantially higher Reduction@P with far lower amortized cost (e.g., \texttt{GPT-5.4-mini} at $14.2$ LinePerFix@P).
Open-source models show the lowest amortized cost (\texttt{Qwen2.5-Coder-0.5B} at $9.5$ LinePerFix@P); however, LinePerFix@P is meaningful only when paired with positive Reduction@P.
Thus, edit minimality is meaningful only as a constrained efficiency measure, not as an independent ranking signal.
Supplementary structural diagnostics likewise indicate that passing revisions are generally localized (Appendix~\ref{sec:appendix_camera_ready}).

\paragraph{Failure Analysis}
For diagnostic analysis only, we analogously compute \texttt{@F} metrics over failed instances $D \setminus S_{\text{pass}}$; these metrics are not used as primary quality metrics.
Data in Table~\ref{tab:failure_analysis} illustrates distinct behavioral patterns during refactoring failures.
For closed-source models, both lint alert reduction and average edit volume remain similar across passed and failed runs, showing smaller divergence than open-source models.
Conversely, several open-source models exhibit a stark asymmetric shift, where lint alert reduction drops from positive values to strongly negative values during failed runs (e.g., \texttt{Qwen2.5-Coder-7B} dropping from $24.7\%$ to $-88.2\%$), correlating with a reduction in average edit volume.
This pattern indicates that on failed instances, open-source models' failures are not driven by excessive rewriting, but by weak target selection: fewer edits still introduce more new alerts than they remove, while also failing functionality preservation.
These two failure modes mark the task frontier: closed-source commercial models face a functionality preservation bottleneck once lint alert reduction is largely achieved, whereas open-source models must first learn safe, targeted edits.
A complementary failure-type breakdown confirms that build success is not a reliable proxy for functionality preservation (Table~\ref{app:cr_failures}).

\subsection{Influencing Factors}

\paragraph{Programming Language}
Table~\ref{tab:language_performance} reveals a pronounced cross-lingual performance gradient across all paradigms.
 C++ emerges as the most stable domain, where even open-source models consistently achieve positive lint alert reduction; this pattern may reflect differences in linter rule sets, alert category distribution, and dataset composition rather than language typing discipline alone.
 Conversely, Python and Java present higher optimization barriers: while closed-source commercial models remain above $77\%$ Reduction@P, some open-source models exhibit negative Reduction@P.
 Notably, \texttt{Qwen2.5-Coder-7B} reaches its lowest Reduction@P on Python ($19.0\%$) and remains low on Java ($21.6\%$), both far below its C++ result ($27.6\%$).
One plausible explanation is that many C++ alerts correspond to localized transformation patterns, whereas Python and Java alerts more often involve naming, convention, or structure-level changes that require broader surrounding context.

\paragraph{Lint Alert Category.}
\begin{figure*}[t]
    \centering
    \includegraphics[width=\linewidth]{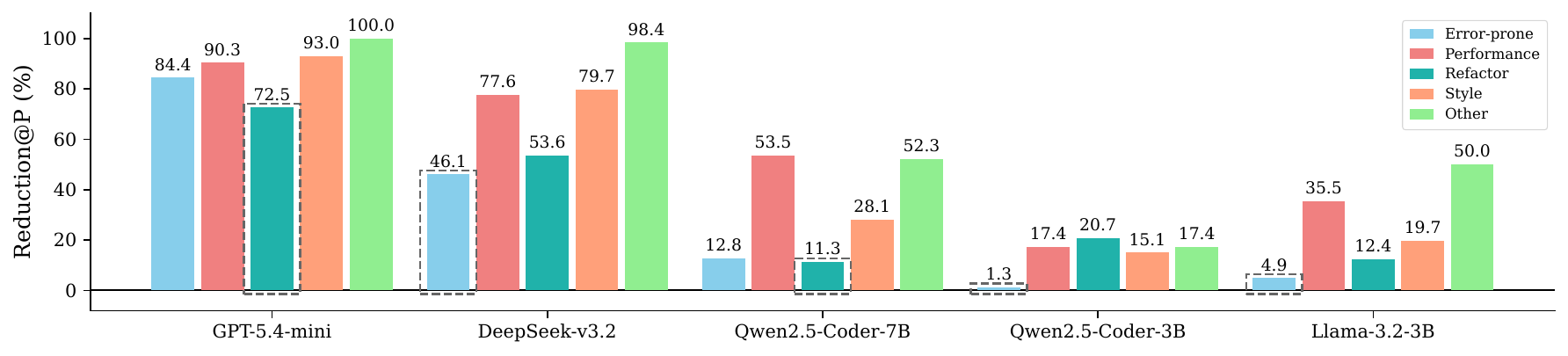}
    \caption{Category-wise Reduction@P for selected models. Additional category-wise results, including negative values for smaller models, are reported in Table~\ref{app:category_performance_full}.}
    \label{fig:category_breakdown}
\end{figure*}

% ============================================================
% Ablation: With-Alerts vs Without-Alerts
% ============================================================
\begin{table*}[t]
\centering
\small
\resizebox{\linewidth}{!}{
\begin{tabular}{lcccccccccc}
\toprule
 & \multicolumn{5}{c}{\textbf{With Alerts}} & \multicolumn{5}{c}{\textbf{No Alerts}} \\
\cmidrule(lr){2-6} \cmidrule(lr){7-11}
\textbf{Model} & Pass@1 & Reduction@P & New@P & $\Delta$Line@P & LinePerFix@P & Pass@1 & Reduction@P & New@P & $\Delta$Line@P & LinePerFix@P \\
\midrule
GPT-5.4-mini & 68.3 & 90.4 & 5.6 & 60.2 & 14.2 & 95.4 \textcolor{green!70!black}{$\uparrow$} & 45.4 \textcolor{red!70!black}{$\downarrow$} & 46.8 \textcolor{red!70!black}{$\uparrow$} & 69.0 \textcolor{red!70!black}{$\uparrow$} & 15.5 \textcolor{red!70!black}{$\uparrow$} \\
DeepSeek-v3.2 & 57.6 & 73.0 & 6.7 & 45.4 & 13.8 & 62.5 \textcolor{green!70!black}{$\uparrow$} & 49.9 \textcolor{red!70!black}{$\downarrow$} & 10.1 \textcolor{red!70!black}{$\uparrow$} & 81.6 \textcolor{red!70!black}{$\uparrow$} & 40.5 \textcolor{red!70!black}{$\uparrow$} \\
Qwen2.5-Coder-0.5B & 48.8 & -1.2 & 10.5 & 17.2 & 9.5 & 60.0 \textcolor{green!70!black}{$\uparrow$} & -1.0 \textcolor{green!70!black}{$\uparrow$} & 3.4 \textcolor{green!70!black}{$\downarrow$} & 10.5 \textcolor{green!70!black}{$\downarrow$} & 74.4 \textcolor{red!70!black}{$\uparrow$} \\
Qwen2.5-Coder-7B & 53.0 & 24.7 & 12.7 & 32.5 & 12.7 & 64.5 \textcolor{green!70!black}{$\uparrow$} & 9.2 \textcolor{red!70!black}{$\downarrow$} & 3.6 \textcolor{green!70!black}{$\downarrow$} & 31.0 \textcolor{green!70!black}{$\downarrow$} & 42.9 \textcolor{red!70!black}{$\uparrow$} \\
\bottomrule
\end{tabular}
}
\caption{Ablation study on the impact of lint alert conditioning. \textit{With Alerts}: full input with lint diagnostics; \textit{No Alerts}: input without lint alert messages. Arrows in the No Alerts columns compare against the With Alerts baseline: \textcolor{green!70!black}{$\uparrow$}/\textcolor{green!70!black}{$\downarrow$} = improvement; \textcolor{red!70!black}{$\uparrow$}/\textcolor{red!70!black}{$\downarrow$} = degradation.}
\label{tab:ablation_alerts}
\end{table*}

Alert category difficulty varies sharply across model scales, with Error-prone and Style alerts posing the greatest challenge (Figure~\ref{fig:category_breakdown}).
GPT-5.4-mini achieves above $84\%$ Reduction@P on Error-prone and above $90\%$ on Performance and Style categories; \texttt{DeepSeek-v3.2} shows a different profile, excelling on Performance ($77.6\%$) and Refactor ($53.6\%$) but dropping to $46.1\%$ on Error-prone.
Smaller open-source models are less reliable overall: the 0.5B model reaches only $-1.2\%$ Reduction@P, while the 7B model reaches $24.7\%$ and preserves substantially more functionality. Nevertheless, even the 7B model remains sensitive to language and category, with only $19.0\%$ Reduction@P on Python and $21.6\%$ on Java (additional size- and category-specific results are reported in Appendix~\ref{sec:appendix_category}, Table~\ref{app:category_performance_full}).
The three categories appear to stress different capabilities: Performance alerts map to localized idioms applicable with little context, whereas Error-prone alerts require semantic judgment that stresses functionality preservation, and Style alerts---the most numerous ($67.3\%$)---are fine-grained, so partial rewrites readily leave residual or inconsistent violations.
Performing well on LintEval thus requires both high-volume cleanup and semantically sensitive modification.
At the instance level, only 36.6\% of programs are Style-only; 43.6\% are mixed and 3.2\% contain potentially behavior-relevant rules only, so at least 46.8\% contain a potentially behavior-relevant rule (Appendix~\ref{sec:appendix_camera_ready}). At the rule level, Style-only rules account for 78.9\% of alerts.

\paragraph{Alert Density.}
Higher alert density amplifies difficulty. For \texttt{GPT-5.4-mini}, Pass@1 drops from $82.1\%$ on single-alert files to $63.9\%$ on files with more than $10$ alerts, while Reduction@P rises from $62.5\%$ to $92.1\%$. Capable models thus resolve clustered alerts in bulk, but the resulting broader modification tendency contributes to the Pass@1 decline of $18.2$ percentage points.
This density effect parallels the refactoring-oriented SFT format (Section~\ref{sec:sft}): in both, more input lint alerts are associated with a broader edit scope, suggesting the pressure toward broader file-level rewriting arises partly from the alert signal itself.
Full scaling curves are provided in Appendix~\ref{sec:appendix_warning} (Table~\ref{app:alert_count_full}).

% ============================================================
% Placement: Main Text (Section 5 - SFT Experiments)
% ============================================================
\begin{table*}[t]
\centering
\small
\begin{tabular}{llccccc}
\toprule
\textbf{Model} & \textbf{Format} & \textbf{Pass@1} & \textbf{Reduction@P} & \textbf{New@P} & \textbf{$\Delta$Line@P} & \textbf{LinePerFix@P} \\
\midrule
\multirow{4}{*}{0.5B} & Base & \textbf{48.8\%} & -1.2\% & 10.5\% & 17.2 & \textbf{9.5} \\
 & A+L$\rightarrow$A$'$ & 37.8\% & \textbf{79.9\%} & 8.4\% & 42.0 & 15.9 \\
 & Q$\rightarrow$A & 45.2\% & -1.0\% & \textbf{2.5\%} & \textbf{2.6} & 30.8 \\
 & Q$\rightarrow$A$'$ & 40.8\% & 9.1\% & 3.7\% & 11.9 & 16.3 \\
\midrule
\multirow{4}{*}{3B} & Base & 52.0\% & 9.5\% & 11.2\% & 27.9 & 12.8 \\
 & A+L$\rightarrow$A$'$ & 60.4\% & \textbf{85.0\%} & 6.6\% & 56.1 & 16.0 \\
 & Q$\rightarrow$A & \textbf{65.1\%} & 6.6\% & \textbf{2.7\%} & \textbf{4.1} & \textbf{9.7} \\
 & Q$\rightarrow$A$'$ & 51.0\% & 10.2\% & 22.2\% & 32.3 & 12.3 \\
\midrule
\multirow{4}{*}{7B} & Base & 53.0\% & 24.7\% & 12.7\% & 32.5 & 12.7 \\
 & A+L$\rightarrow$A$'$ & 61.2\% & \textbf{76.0\%} & 7.0\% & 49.6 & 16.0 \\
 & Q$\rightarrow$A & \textbf{68.3\%} & 15.8\% & \textbf{4.7\%} & \textbf{12.4} & \textbf{10.0} \\
 & Q$\rightarrow$A$'$ & 66.7\% & 47.9\% & 4.9\% & 33.5 & 14.4 \\
\bottomrule
\end{tabular}
\caption{Effect of training data format on \textbf{LintEval}. Three Qwen2.5-Coder sizes (0.5B, 3B, 7B) are fine-tuned with formats A+L$\rightarrow$A$'$, Q$\rightarrow$A, and Q$\rightarrow$A$'$. \textbf{Bold} marks the best value per model group.}
\label{tab:sft_results}
\end{table*}

\paragraph{Linter Guidance (With vs.\ Without Alerts).}
To isolate the contribution of explicit lint alerts, we re-run all models with an alert-free prompt that retains the refactoring instruction but omits the lint alert list.
Removing alerts improves Pass@1 by $4.9$--$27.1$ percentage points, but weakens lint alert reduction for models that benefit from alert conditioning and increases LinePerFix@P for all models, with the largest raw increase on \texttt{Qwen2.5-Coder-0.5B} ($9.5$ to $74.4$; Table~\ref{tab:ablation_alerts}).
This pattern suggests that, without lint alerts, models either make conservative edits that preserve functionality but fix few reported violations, or perform untargeted rewrites with poor edit efficiency.
Explicit lint alerts therefore serve as localization signals and intervention signals: they tell models where to edit and what to optimize, but they also increase pressure to modify code that already satisfies functionality preservation.
In a matched Qwen2.5-Coder-7B comparison, combining developer instructions with lint alerts preserves the lint-only Pass@1 ($90.0\%$) and nearly preserves Reduction@P ($39.0\%$ vs. $39.9\%$), while lowering New@P from $10.6\%$ to $7.1\%$; instructions alone reach higher Pass@1 ($95.0\%$) but lower Reduction@P ($25.0\%$) (Appendix~\ref{sec:appendix_camera_ready}). This synthetic-instruction evaluation set is separate from the multi-round set, so their absolute values are not directly comparable.

\paragraph{Iterative Feedback.}
The main protocol measures the controlled first-pass refactoring step. A fixed three-round mini-swe-agent diagnostic (Table~\ref{app:cr_multiround}) shows that conditional Reduction@P increases from Round 1 to Round 3 by 9.0, 10.0, and 1.9 points for GPT-5.4-mini, DeepSeek-v3.2, and Qwen2.5-Coder-7B, while Pass@1 falls by 1.6, 3.8, and 1.2 points. Because each @P value conditions on that round's passing outputs, these differences are descriptive rather than causal effects of feedback.

\subsection{Impact of Instruction Tuning}
\label{sec:sft}

\paragraph{Data and Training Setup.}
We construct the fine-tuning corpus from 1,196 problem instances outside LintEval to avoid problem-level contamination. For each instance we collect the problem statement $Q$, the original implementation $A$, its lint alerts $L$, and a refined implementation $A'$ distilled from \texttt{GPT-5.2}. A $10\%$ stratified sample is manually audited to confirm that distilled revisions remove the intended alerts while preserving behavior; full corpus statistics are provided in Appendix~\ref{sec:appendix_sft_data} (Table~\ref{app:sft_data_statistics}). We fine-tune the Qwen2.5-Coder family with LoRA \citep{hu2021lora} ($r{=}16$, $\alpha{=}32$, learning rate $2\times 10^{-5}$, effective batch size $8$, $3$ epochs) under three formulations: \textbf{A+L$\to$A$'$} (refactoring-oriented: original code plus alerts $\to$ refined code), \textbf{Q$\to$A} (vanilla generation: problem $\to$ original code), and \textbf{Q$\to$A$'$} (quality-purified generation: problem $\to$ refined code).

\paragraph{Refactoring-oriented supervision.}
\textbf{A+L$\to$A$'$} provides the closest match to linter-guided file refactoring because the model observes both the original file and its lint alerts during training.
This alignment yields the strongest lint alert reduction, with the \texttt{3B} model improving from $9.5\%$ to $85.0\%$ Reduction@P.
However, the same format also produces the largest edit scope across all scales, suggesting that it learns to resolve lint alerts through broader file-level rewrites rather than minimal local changes.
This helps explain why \textbf{A+L$\to$A$'$} improves new alert avoidance relative to Base but still has higher New@P and lower Pass@1 than \textbf{Q$\to$A}.
Refactoring-oriented supervision is therefore effective for lint alert reduction, but it does not by itself guarantee functionality preservation or edit minimality.
This scope dependence is scale-sensitive: \textbf{A+L$\to$A$'$} raises Pass@1 at 3B and 7B but lowers it at 0.5B ($48.8\%$$\to$$37.8\%$), suggesting the format's multi-source conditioning becomes reliable only above a minimum model capacity.
The direction of the Reduction@P effect is stable across language and size: A+L$\to$A$'$ exceeds Base in all nine language-size cells, with gains of $+23.0$ to $+89.6$ points, while Pass@1 changes are mixed, ranging from $-22.5$ to $+23.6$ points (Table~\ref{app:cr_language_sft}). Additional checks extend the comparison beyond one training regime: full-parameter SFT on Qwen2.5-Coder-3B improves A+L$\to$A$'$ from 59.0\% to 63.7\% Pass@1 and from 84.0\% to 85.8\% Reduction@P. On DeepSeek-Coder-6.7B-Instruct, A+L$\to$A$'$ improves over Base from 25.3\% to 61.2\% Pass@1 and from 19.0\% to 79.0\% Reduction@P, and is the only tested format above Base on both metrics (Table~\ref{app:cr_sft_generalization}).

\paragraph{Generation-oriented supervision.}
\textbf{Q$\to$A} optimizes a different behavior: generating an original correct solution from the problem statement rather than modifying a given file under lint alerts.
This format yields the strongest functionality preservation among the fine-tuned variants, reaching $68.3\%$ Pass@1 on \texttt{7B}, and consistently gives the lowest New@P.
The cost is limited lint alert reduction: for \texttt{7B}, Reduction@P remains $15.8\%$, below the Base result of $24.7\%$.
This pattern suggests that generation-oriented supervision encourages conservative edits and new alert avoidance, but does not teach models to use lint alerts as actionable refactoring targets.

\paragraph{Quality-purified generation.}
\textbf{Q$\to$A$'$} exposes models to refined implementations, but it still omits the original file and its lint alerts from the training input.
As a result, it can transfer a code-quality prior without learning the conditional editing behavior required by linter-guided file refactoring.
This explains its mixed behavior: on \texttt{7B}, it improves Reduction@P to $47.9\%$ while keeping Pass@1 close to \textbf{Q$\to$A}; on \texttt{3B}, however, New@P rises to $22.2\%$, indicating unstable new alert avoidance.
Quality-purified generation is therefore useful only when the model can translate the cleaner target distribution into controlled file-level edits.

\section{Related Work}
\label{sec:related_work}

\paragraph{Bug Repair Benchmarks}
Automated program repair benchmarks evaluate debugging proficiency across standalone snippets or complex repository-level environments, using real-world faults such as Defects4J \citep{just2014defects4j}, debugging tasks in DebugBench \citep{tian2024debugbench}, repository-level issues in SWE-bench \citep{jimenez2023swe}, and learned detect-localize-repair pipelines such as CodeT5 \citep{bui2022detect}. Interactive code-generation and execution-feedback methods, including Reflexion, InterCode, and FeedbackEval \citep{shinn2023reflexion, yang2023intercode, dai2025feedbackeval, huang2022execution}, study related feedback-driven behavior but are not specifically bug-repair benchmarks.
These works primarily target debugging or code generation, rather than improving code quality under strict functionality preservation, which motivates a dedicated benchmark.

\paragraph{Benchmarks Related to Linter-Guided File Refactoring}
Existing refactoring benchmarks target programmatic modifications but differ significantly in tool guidance and evaluation scope.
RefactorBench \citep{gautam2025refactorbench} measures instruction-driven refactoring under free-form text requests without structured tool feedback.
Concurrently, MetaLint \citep{naik2025metalint} evaluates model capabilities in detecting and localizing best-practice violations from natural language specifications, while LintLLM \citep{fang2025lintllm} targets hardware defect identification in Verilog RTL code.
However, these benchmarks remain domain-specific and focus on passive alert detection and localization rather than active transformation execution under structured lint alerts.

\paragraph{Static Analysis}
Linters and static analyzers are widely used to surface code-quality issues and support maintenance workflows \citep{beller2016analyzing, christakis2016what, tomasdottir2020adoption, zampetti2017how, koszo2025large}. Automated techniques have been developed to address such issues, including template- and program-transformation-based fixes \citep{MarcilioFBP20}, while more recent work uses LLMs together with static-analysis or code-quality feedback \citep{wadhwa2024core, li2024enhancing, li2025bitsaifix, scott2025static, yao2025training}. These paradigms lack a standardized benchmark that jointly evaluates functionality preservation, lint alert reduction, new alert avoidance, and edit minimality—the gap LintEval fills.

\section{Conclusion}
\label{sec:conclusion}

We introduce \textbf{LintEval}, a large-scale, multilingual benchmark for linter-guided file refactoring on programs with correct initial functionality. Under high-coverage test oracles, no method dominates: rule-based baselines preserve functionality but reduce few lint alerts, whereas closed-source commercial models reduce more lint alerts at the cost of functionality preservation, and open-source models show cross-language and alert-category weaknesses partly mitigated by refactoring-oriented fine-tuning. Jointly optimizing these four dimensions remains an open challenge for automated code maintenance.

% Limitations and Ethics: not counted in page limit

\section*{Limitations}
\label{sec:limitations}

Our benchmark primarily evaluates linter-guided file refactoring within an explicitly controlled environment using algorithmic tasks from \textit{CodeContests+}.
While this setup provides high-coverage test oracles to help isolate core capabilities, it omits production-level architectural complexities such as cross-file dependencies or multi-turn iterative developer workflows.
Additionally, our evaluation measures dynamic functionality preservation via Pass@1 metrics rather than formal semantic equivalence or structural maintainability metrics.
Finally, our supervised fine-tuning analysis is bounded in scale and optimized primarily for the \texttt{Qwen2.5-Coder} architecture, which may introduce architectural biases across specific lint alert categories and open-source models.

\section*{Ethical Considerations}

This work follows the ACL Ethics Policy.
We utilize the publicly available \textit{CodeContests+} dataset under its original open-source licensing, ensuring that no personal, private, or sensitive user data is involved in our benchmark construction.
We used generative AI writing assistance only for language polishing and grammatical refinement; all experimental design, analysis, and claims were produced and verified by the authors.

\section*{Acknowledgments}
We thank the anonymous reviewers for their constructive feedback.
This work was supported by the National Key Research and Development Program of China under Grant No. 2023YFB4503803.

\bibliography{custom}

\clearpage
\appendix

\section*{Appendix}
\section{Reproducibility Details and Additional Results}
\label{sec:appendix}

This appendix provides reproducibility details, extended quantitative results, and qualitative examples that complement the main text. All central claims, methods, and conclusions are presented in the main body; the appendix records additional information useful for replication and closer inspection.

% --- Put the camera-ready diagnostics first so their floats are resolved early ---

\subsection{Additional Diagnostic Results}
\label{sec:appendix_camera_ready}

These diagnostics answer four questions raised during review: whether iterative feedback changes the functionality--reduction trade-off, whether the findings persist for newer backbones, what causes functional failures, and whether the SFT effect transfers across languages and training settings. We report full tables only when row-level comparisons matter; compact audits are summarized in prose below.

Tables 9--12 present four complementary diagnostic analyses with different units and comparison scopes. The iterative-feedback table compares two rounds of the same procedure, the failure table reports percentages within non-passing revisions, and the two SFT tables report either differences from Base or absolute performance under a training setting. Consequently, values should be compared within each table rather than combined across tables. In particular, conditional Reduction@P always describes the subset that passes functionality tests in the corresponding round or setting; it is not an unconditional estimate over all attempted revisions.

\paragraph{Iterative feedback and model recency.}
Three rounds of a fixed mini-swe-agent loop increase conditional Reduction@P for all three tested models, but Pass@1 decreases slightly (Table~\ref{app:cr_multiround}). Thus, repeated linter feedback helps remove alerts without resolving the functionality trade-off. The newer-Qwen comparison in the main paper (Table~\ref{tab:new_qwen_check}) reaches the same conclusion from a different direction: Qwen3-8B has the highest Reduction@P but the lowest Pass@1, while Qwen3.6-35B-A3B has the highest Pass@1 but lower Reduction@P. No tested checkpoint dominates both dimensions.

The magnitude of the multi-round gain is also model-dependent. DeepSeek-v3.2 obtains the largest Reduction@P increase (+10.0 points), closely followed by GPT-5.4-mini (+9.0), whereas \texttt{Qwen2.5-Coder-7B} improves by only 1.9 points. All three incur a Pass@1 loss, ranging from 1.2 to 3.8 points. This directional consistency suggests that another round primarily creates additional opportunities to edit remaining alerts; it does not act as a general correction mechanism for behavioral regressions. Since the passing subset can change between rounds, these differences characterize the observed trajectory of the fixed loop and should not be interpreted as a causal gain on an identical set of successful outputs.

% Diagnostics committed during rebuttal and incorporated in the camera-ready version.
\begin{table}[t]
\centering
\small
\setlength{\tabcolsep}{3pt}
\begin{tabular}{lccc}
\toprule
\textbf{Model} & \makecell{Round 1\\P / R} & \makecell{Round 3\\P / R} & \makecell{Change\\R / P} \\
\midrule
GPT-5.4-mini & 67.5 / 77.9 & 65.9 / 86.9 & +9.0 / -1.6 \\
DeepSeek-v3.2 & 67.1 / 71.3 & 63.3 / 81.3 & +10.0 / -3.8 \\
Qwen2.5-Coder-7B & 56.4 / 18.7 & 55.2 / 20.6 & +1.9 / -1.2 \\
\bottomrule
\end{tabular}
\caption{Three-round mini-swe-agent diagnostic on the same LintEval instances. P and R denote Pass@1 and Reduction@P. Each Reduction@P conditions on that round's passing outputs.}
\label{app:cr_multiround}
\end{table}

\begin{table}[t]
\centering
\small
\setlength{\tabcolsep}{1.2pt}
\begin{tabular}{lrrrrrr}
\toprule
\textbf{Model} & Build & Run & TO & Wrong & $\geq$90\% & 0\% \\
\midrule
GPT-5.4-mini & 10.5 & 5.3 & 1.8 & 82.5 & 19.1 & 2.1 \\
DeepSeek-v3.2 & 28.1 & 6.2 & 1.6 & 64.1 & 22.0 & 4.9 \\
Qwen2.5-Coder-7B & 26.3 & 13.2 & 1.3 & 59.2 & 17.8 & 6.7 \\
Qwen2.5-Coder-0.5B & 40.8 & 9.7 & 0.0 & 49.5 & 15.7 & 13.7 \\
\bottomrule
\end{tabular}
\caption{Camera-ready diagnostics (continued): failure types among non-passing revisions (\%). Run and TO denote runtime errors and timeouts; the last two columns split wrong answers by test pass rate. Percentages may not sum to 100\% because of rounding.}
\label{app:cr_failures}
\end{table}

\begin{table}[t]
\centering
\small
\setlength{\tabcolsep}{1pt}
\begin{tabular}{llrr}
\toprule
\textbf{Size} & \textbf{Language} & \makecell{Pass@1\\diff.} & \makecell{Reduction@P\\diff.} \\
\midrule
Qwen2.5-Coder-0.5B & C++ & -22.5 & +89.6 \\
Qwen2.5-Coder-0.5B & Python & +1.5 & +78.9 \\
Qwen2.5-Coder-0.5B & Java & -5.9 & +56.6 \\
Qwen2.5-Coder-3B & C++ & -4.5 & +81.2 \\
Qwen2.5-Coder-3B & Python & +23.1 & +73.7 \\
Qwen2.5-Coder-3B & Java & +11.8 & +63.3 \\
Qwen2.5-Coder-7B & C++ & -2.4 & +56.6 \\
Qwen2.5-Coder-7B & Python & +23.6 & +53.7 \\
Qwen2.5-Coder-7B & Java & -1.9 & +23.0 \\
\bottomrule
\end{tabular}
\caption{Camera-ready diagnostics (continued): language--size effects of A+L$\to$A$'$ supervision relative to Base. Reduction@P improves in all nine cells, while Pass@1 changes are mixed.}
\label{app:cr_language_sft}
\end{table}

\begin{table}[t]
\centering
\small
\setlength{\tabcolsep}{1.5pt}
\begin{tabular}{lrr}
\toprule
\textbf{Setting} & Pass@1 & Reduction@P \\
\midrule
Qwen2.5-Coder-3B LoRA & 59.0 & 84.0 \\
Qwen2.5-Coder-3B full-parameter & 63.7 & 85.8 \\
DeepSeek-Coder-6.7B Base & 25.3 & 19.0 \\
DeepSeek-Coder-6.7B A+L$\to$A$'$ & 61.2 & 79.0 \\
DeepSeek-Coder-6.7B Q$\to$A & 24.9 & 13.0 \\
DeepSeek-Coder-6.7B Q$\to$A$'$ & 8.6 & 18.4 \\
\bottomrule
\end{tabular}
\caption{Camera-ready diagnostics (continued): SFT robustness across training regimes and backbones. Reduction@P conditions on each setting's passing outputs.}
\label{app:cr_sft_generalization}
\end{table}

\paragraph{Failure modes and tool coverage.}
Wrong answers dominate non-passing revisions across model families (Table~\ref{app:cr_failures}); only 15.7\%--22.0\% of wrong-answer cases pass at least 90\% of tests, so successful compilation is not a reliable proxy for functionality. The rule-based coverage audit provides a separate explanation for its limited alert reduction: the selected tool/analyzer pairings fix 77.0\% of C++ alerts, 36.8\% of Python alerts, and 5.2\% of Java alerts, retaining 23.0\%, 63.2\%, and 94.8\%, respectively. These shares use all original alerts and are not an arithmetic decomposition of conditional Reduction@P.

The balance among non-semantic failures varies by model. Build failures account for 10.5\% of GPT-5.4-mini failures but 40.8\% for \texttt{Qwen2.5-Coder-0.5B}; runtime errors likewise rise from 5.3\% to 13.2\% for the tested models with the highest observed rate. Nevertheless, wrong answers remain the plurality or majority in every row. This distinction matters operationally: syntax repair, compilation feedback, or sandbox retries can address build and runtime failures, but they cannot establish semantic equivalence for revisions that execute and still return incorrect results. The failure table therefore motivates retaining execution-based functionality tests as the primary gate rather than replacing them with build success.

\paragraph{Alert composition and analyzer dependence.}
At the instance level, 36.6\% of programs contain Style-only rules, 43.6\% mix Style with potentially behavior-relevant rules, 3.2\% contain only potentially behavior-relevant rules, and 16.7\% are unclear. At the rule level, the corresponding shares are 78.9\%, 17.9\%, and 3.3\% (unclear). These labels characterize rule categories, not semantic-equivalence outcomes. A matched additional-analyzer audit finds files flagged only by Cppcheck, Ruff analysis mode, and Checkstyle in 16.0\%, 32.0\%, and 49.0\% of C++, Python, and Java cases, respectively; this establishes analyzer dependence but does not imply transfer of model rankings.

% Language-specific linter configurations and runtime environments.
\begin{table*}[t]
\centering
\small
\begin{tabular}{lcccc}
\toprule
\textbf{Language} & \textbf{Linter} & \textbf{Rule Family} & \textbf{Linter Version} & \textbf{Compiler/Runtime} \\
\midrule
C++ & clang-tidy & bugprone, performance, modernize, readability & 22.1.0 & g++ 13.3.0 \\
Python & Pylint & error, performance, refactor, convention & 4.0.5 & Python 3.12.10 \\
Java & PMD & Error Prone, Performance, Design, Code Style & 7.12.0 & OpenJDK 21.0.10 \\
\bottomrule
\end{tabular}

\caption{Language-specific linter configurations and runtime environments.}
\label{app:execution_env}
\end{table*}

% ============================================================
% Placement: Main Text (Section 2 - Taxonomy)
% ============================================================
\begin{table*}[t]
\centering
\small
\begin{tabular}{lc lll}
\toprule
\textbf{Category} & \textbf{Lint Alerts} & \textbf{C++ (clang-tidy)} & \textbf{Python (Pylint)} & \textbf{Java (PMD)} \\
\midrule
Error-prone & 18,434  & \texttt{bugprone-*}    & \texttt{error}      & \texttt{Error Prone}, \texttt{Security} \\
Performance & 17,776  & \texttt{performance-*} & \texttt{performance}& \texttt{Performance}, \texttt{Multithreading} \\
Refactor    & 17,977  & \texttt{modernize-*}   & \texttt{refactor}   & \texttt{Design} \\
Style       & 115,458 & \texttt{readability-*} & \texttt{convention} & \texttt{Code Style}, \texttt{Best Practices} \\
Other       & 2,017   & \texttt{other}         & \texttt{other}      & \texttt{other} \\
\bottomrule
\end{tabular}
\caption{Cross-language taxonomy of lint alert categories in LintEval. Counts show the number of alerts in each diagnostic category. Representative rule families illustrate how each category maps to language-specific linter conventions.}
\label{tab:cross_language_mapping}
\end{table*}

\paragraph{SFT robustness and instruction source.}

Refactoring-oriented A+L$\to$A$'$ supervision improves Reduction@P over Base in all nine language--size cells, while Pass@1 changes are mixed (Table~\ref{app:cr_language_sft}). The effect also survives full-parameter Qwen2.5-Coder-3B training and a DeepSeek-Coder-6.7B backbone (Table~\ref{app:cr_sft_generalization}). In a matched Qwen2.5-Coder-7B input audit, lint alerts achieve 90.0\% Pass@1, 39.9\% Reduction@P, and 10.6\% New@P; developer instructions alone yield 95.0\%, 25.0\%, and 6.0\%; combining both yields 90.0\%, 39.0\%, and 7.1\%. The combined input therefore retains nearly all alert reduction while introducing fewer new alerts.

The language--size table separates robustness of the optimization target from robustness of functionality preservation. Reduction@P gains are positive in every cell, from +23.0 points for 7B Java to +89.6 points for 0.5B C++, showing that the supervision format consistently teaches alert removal. Pass@1, however, improves mainly on Python at 3B and 7B and on Java at 3B, while decreasing in the remaining cells. Full-parameter training does not reverse this pattern: it modestly improves both reported metrics over the matched 3B LoRA setting, but it does not by itself demonstrate that broader updates eliminate the trade-off.

The DeepSeek-Coder results provide a stronger transfer check because both the backbone and model size change. A+L$\to$A$'$ raises Pass@1 from 25.3\% to 61.2\% and Reduction@P from 19.0\% to 79.0\%, whereas the two generation-oriented formats do not improve both metrics over Base. Together, these diagnostics support a bounded conclusion: refactoring-oriented supervision transfers as an alert-reduction signal across the tested languages, scales, training regimes, and backbones, but its functionality effect remains contingent on model capacity and setting.

\paragraph{Diagnostic conclusion.}
Across the four tables, additional computation and targeted supervision both increase alert reduction, yet only the refactoring-oriented transfer check improves functionality and reduction together. Iteration alone slightly lowers Pass@1, and failure decomposition shows that many regressions are semantic rather than merely syntactic. The diagnostics therefore reinforce the main evaluation design: alert reduction must remain conditioned on independently verified functionality, and improvements should be reported jointly rather than collapsed into a single cleanup score.

\subsection{Cross-Language Taxonomy}
\label{sec:appendix_taxonomy}

Table~\ref{tab:cross_language_mapping} presents the cross-language taxonomy of lint alert categories used in LintEval. The taxonomy groups heterogeneous linter outputs from clang-tidy, Pylint, and PMD into five unified categories: Error-prone, Performance, Refactor, Style, and Other. This normalization enables aggregate cross-language analysis while preserving the original rule identifiers for per-rule inspection.

Style alerts dominate the dataset (67.3\%), while Error-prone, Performance, and Refactor alerts collectively account for 31.6\%, making consistent file-wide editing a central challenge. Our taxonomy groups rules by semantic concern rather than rule-name similarity; the released data retains each source analyzer and rule identifier so the aggregate categories remain auditable and can be regrouped for analyzer-specific analysis.

% ============================================================
% Placement: Appendix (Experimental Setup reproducibility)
% ============================================================
\begin{table}[t]
\centering
\small
\begin{adjustbox}{max width=\columnwidth}
\begin{tabular}{llccc}
\toprule
\textbf{Model} & \textbf{Access} & \textbf{Temp.} & \textbf{Top-$p$} & \textbf{Max Tokens} \\
\midrule
GPT-5.2 & API & 0.0 & 1.0 & 4,096 \\
GPT-5.4-mini & API & 0.0 & 1.0 & 4,096 \\
\midrule
DeepSeek-v3.2 & API & 0.0 & 1.0 & 4,096 \\
Qwen2.5-Coder-32B & local & 0.0 & 1.0 & 4,096 \\
Qwen2.5-Coder-7B & local & 0.0 & 1.0 & 4,096 \\
Qwen2.5-Coder-3B & local & 0.0 & 1.0 & 4,096 \\
Qwen2.5-Coder-0.5B & local & 0.0 & 1.0 & 4,096 \\
Llama-3.2-3B & local & 0.0 & 1.0 & 4,096 \\
Llama-3.2-1B & local & 0.0 & 1.0 & 4,096 \\
\bottomrule
\end{tabular}
\end{adjustbox}
\caption{Decoding configurations for all evaluated models. All models use greedy decoding. Local models are served via vLLM.}
\label{app:model_decoding_config}
\end{table}

\subsection{Execution Environment}
\label{sec:appendix_execution_env}

Table~\ref{app:execution_env} reports the linter versions and runtime configurations used for lint alert extraction. Reproducibility of the linting stage is critical because linter diagnostics can vary across versions, enabled rule sets, and runtime environments. For example, clang-tidy 22.1.0 introduces several new modernize-* checks compared to earlier versions, and PMD 7.12.0 reorganizes rule categories differently from PMD 6.x.

Across all three languages, each execution is subject to a 10-second timeout and a 1GB memory limit to prevent runaway analysis on pathological inputs. The compilers and runtime versions (g++ 13.3.0, Python 3.12.10, OpenJDK 21.0.10) are pinned to ensure that functionality tests execute consistently across runs.

These resource limits apply uniformly across models and languages; they are part of the evaluation protocol rather than model-specific tuning. Pinning both analyzer and execution environments is necessary because the benchmark measures two linked outcomes: alerts are counted under the analyzer configuration, while functionality is decided by language-specific execution. Changing either side can alter the passing subset and therefore every conditional @P metric. Reproduction should consequently treat the versions, enabled checks, timeout, and memory cap as a single configuration rather than substituting tools independently.

\subsection{Prompt Template}
\label{sec:appendix_prompt}

The full prompt template used for evaluation is shown below. The prompt is designed to mirror realistic tool-feedback workflows where developers receive linter output and must revise code accordingly.

\begin{figure}[t]
\centering
\begin{tcolorbox}[colback=gray!5,colframe=gray!50,sharp corners,title={Linter-Guided File Refactoring Prompt}]
\ttfamily\small
You are a code quality assistant. Your task is to revise the given \{language\} code to resolve the reported lint alerts while strictly preserving its original functionality.

\textbf{Original Code:}
\{code\}

\textbf{Lint Alerts:}
\{messages\}

\textbf{Requirements:}
1. Preserve the original functionality. The revised code must pass all existing tests.
2. Address as many lint alerts as possible.
3. Do NOT introduce new lint alerts.
4. Make minimal, targeted edits. Avoid unnecessary rewrites.
5. Output ONLY the revised code inside a single code block. No explanations.

\textbf{Revised Code:}
\end{tcolorbox}
\end{figure}

\noindent The prompt provides five key elements: (1) the programming language to set context, (2) the original accepted code as the starting point, (3) the full list of lint alerts with rule IDs, file paths, and line numbers as the modification signal, (4) explicit requirements for functionality preservation, maximal lint reduction, minimal new-alert introduction, and minimal editing, and (5) a code-block-only output constraint to facilitate automated extraction.

\subsection{Model Decoding Configuration}
\label{sec:appendix_decoding}

Table~\ref{app:model_decoding_config} lists the decoding parameters for all evaluated models. All models use greedy decoding (temperature $=0.0$) with the same prompt template. This standardized setup ensures that performance differences are attributable to model behavior rather than sampling variation or prompt engineering.

Greedy decoding with temperature $=0.0$ provides deterministic, reproducible outputs and makes Pass@1 a measure of single-attempt performance. This policy keeps the benchmark focused on conditional editing rather than candidate search and reranking: each model receives the same information and produces one revision, while multi-round or test-guided selection systems require separate evaluation because they use additional attempts and feedback.

\subsection{Full Language-wise Performance Results}
\label{sec:appendix_language_full}

% ============================================================
% Placement: Appendix (Section A - Full Results)
% ============================================================
% A1: Full Language-wise Performance (C++ and Python)
\begin{table*}[t]
\centering
\small
\setlength{\tabcolsep}{2pt}
\begin{tabular}{lcccccccc}
\toprule
& \multicolumn{4}{c}{\textbf{C++}} & \multicolumn{4}{c}{\textbf{Python}} \\
\cmidrule(lr){2-5} \cmidrule(lr){6-9}
\textbf{Model} & Pass@1 & Reduction@P & New@P & LinePerFix@P & Pass@1 & Reduction@P & New@P & LinePerFix@P \\
\midrule
GPT-5.2 & 67.2\% & 94.9\% & 1.3\% & 23.4 & 64.5\% & 84.4\% & 9.7\% & 11.6 \\
GPT-5.4-mini & \textbf{69.7\%} & \textbf{95.6\%} & 0.9\% & 20.7 & \textbf{65.5\%} & \textbf{86.5\%} & \textbf{8.4\%} & \textbf{9.8} \\
\midrule
DeepSeek-v3.2 & 68.0\% & 76.6\% & 1.7\% & 10.7 & 46.8\% & 70.6\% & 9.7\% & 14.5 \\
Qwen2.5-Coder-32B & 64.3\% & 58.6\% & 2.1\% & 12.9 & 58.1\% & 63.5\% & 16.2\% & 10.1 \\
Qwen2.5-Coder-0.5B & 63.9\% & 1.5\% & \textbf{0.1\%} & \textbf{3.2} & 32.5\% & -7.2\% & 28.2\% & 14.9 \\
Qwen2.5-Coder-3B & 66.8\% & 9.5\% & 0.5\% & 9.7 & 35.0\% & 10.1\% & 31.1\% & 15.7 \\
Qwen2.5-Coder-7B & 67.6\% & 27.6\% & 1.2\% & 7.6 & 35.5\% & 19.0\% & 25.3\% & 14.0 \\
Llama-3.2-1B & 44.7\% & 5.2\% & 0.3\% & 4.5 & 23.2\% & -23.6\% & 48.5\% & 21.0 \\
Llama-3.2-3B & 50.8\% & 18.8\% & 1.8\% & 7.0 & 23.6\% & 13.2\% & 47.7\% & 18.6 \\
\bottomrule
\end{tabular}
\caption{Full language-wise performance on LintEval for C++ and Python. All four metrics are reported; best results per column are in \textbf{bold}.}
\label{app:language_full}
\end{table*}

% Full Language-wise Performance (Java)
\begin{table*}[t]
\centering
\small
\setlength{\tabcolsep}{6pt}
\begin{tabular}{lcccc}
\toprule
\textbf{Model} & \multicolumn{1}{c}{Pass@1} & \multicolumn{1}{c}{Reduction@P} & \multicolumn{1}{c}{New@P} & \multicolumn{1}{c}{LinePerFix@P} \\
\midrule
GPT-5.2 & \textbf{76.5\%} & 77.5\% & 17.1\% & 18.8 \\
GPT-5.4-mini & 72.5\% & \textbf{81.7\%} & 15.1\% & \textbf{13.4} \\
\midrule
DeepSeek-v3.2 & 51.0\% & 63.7\% & 20.5\% & 20.0 \\
Qwen2.5-Coder-32B & 62.7\% & 37.2\% & \textbf{14.7\%} & 22.1 \\
Qwen2.5-Coder-0.5B & 41.2\% & -3.6\% & 31.3\% & 16.7 \\
Qwen2.5-Coder-3B & 49.0\% & 8.5\% & 29.6\% & 15.3 \\
Qwen2.5-Coder-7B & 52.9\% & 21.6\% & 42.4\% & 21.4 \\
Llama-3.2-1B & 23.5\% & 18.5\% & 27.2\% & 28.1 \\
Llama-3.2-3B & 41.2\% & 31.4\% & 28.9\% & 17.5 \\
\bottomrule
\end{tabular}
\caption{Full language-wise performance on LintEval (continued): Java. All four metrics are reported; best results per column are in \textbf{bold}.}
\label{app:language_java}
\end{table*}

Tables~\ref{app:language_full} and~\ref{app:language_java} present the complete language-wise performance breakdown with all four metrics for all evaluated models. The first table groups C++ and Python to support a direct comparison between the highest- and lowest-performing language settings, while the second isolates Java so that its New@P and edit-cost patterns remain readable without shrinking the columns. Pass@1 measures functionality preservation; the remaining metrics are conditioned on passing revisions, so they characterize the quality and efficiency of functionally valid edits rather than all generated outputs.

\paragraph{Language-specific difficulty.} Python presents the greatest challenge for open-source models. For example, \texttt{Qwen2.5-Coder-7B} achieves 67.6\% Pass@1 on C++ but only 35.5\% on Python, a gap of 32.1 percentage points. The same model shows 27.6\% Reduction@P on C++ versus 19.0\% on Python. This disparity likely reflects two factors: (1) Python's dynamic typing and flexible syntax make semantic preservation harder to verify through static analysis, and (2) Pylint's rule set includes many convention checks that require understanding of Python idioms (e.g., naming conventions, docstring requirements) which may not be well-represented in code generation training data.

\paragraph{New-alert introduction patterns.} Closed-source models (GPT-5.2, GPT-5.4-mini) maintain low New@P across all languages (0.9\%--17.1\%), while open-source models show higher variance. On Java, \texttt{Qwen2.5-Coder-7B} shows the highest New@P among the evaluated models at 42.4\%, suggesting that while the model can reduce the original alerts, its edits frequently violate other PMD rules. This may reflect the model's weaker familiarity with Java-specific conventions (e.g., exception handling patterns, access modifier usage) compared to C++ and Python.

\paragraph{Edit efficiency trade-offs.} LinePerFix@P reveals a clear trade-off between lint reduction and edit minimality. GPT-5.4-mini achieves the lowest LinePerFix@P on Python (9.8) and Java (13.4), indicating efficient targeted edits. In contrast, \texttt{Llama-3.2-1B} shows LinePerFix@P of 28.1 on Java, suggesting that even when it fixes alerts, the edits are large and potentially disruptive.

\paragraph{Conclusion.} The language split confirms that aggregate performance is not representative of every analyzer--language setting. Closed-source commercial models retain strong alert reduction across all three languages, whereas open-source models lose substantially more ground on Python and Java and often introduce more new alerts there. C++ therefore provides the most stable setting in this benchmark, while robust Python and Java refactoring remains the clearest cross-language bottleneck.

\subsection{Category-wise Performance}
\label{sec:appendix_category}

% ============================================================
% Table: Full Category-wise Performance (Error-prone, Performance, Refactor)
% ============================================================
\begin{table*}[t]
\centering
\small
\setlength{\tabcolsep}{2pt}
\begin{tabular}{lccccccccc}
\toprule
 & \multicolumn{3}{c}{\textbf{Error-prone}} & \multicolumn{3}{c}{\textbf{Performance}} & \multicolumn{3}{c}{\textbf{Refactor}} \\
\cmidrule(lr){2-4} \cmidrule(lr){5-7} \cmidrule(lr){8-10}
\textbf{Model} & Pass@1 & Reduction@P & New@P & Pass@1 & Reduction@P & New@P & Pass@1 & Reduction@P & New@P \\
\midrule
GPT-5.2 & 63.3\% & 83.0\% & 0.7\% & 65.8\% & 88.9\% & 1.3\% & \textbf{62.0\%} & 59.3\% & 10.5\% \\
GPT-5.4-mini & \textbf{64.9\%} & \textbf{84.4\%} & \textbf{0.0\%} & \textbf{66.7\%} & \textbf{90.3\%} & \textbf{0.2\%} & 60.8\% & \textbf{72.5\%} & 11.0\% \\
\midrule
DeepSeek-v3.2 & 59.6\% & 46.1\% & \textbf{0.0\%} & 53.9\% & 77.6\% & 0.7\% & 53.0\% & 53.6\% & 10.7\% \\
Qwen2.5-Coder-32B & 61.3\% & 33.3\% & 1.3\% & 60.2\% & 73.5\% & 4.2\% & 51.9\% & 55.6\% & 3.7\% \\
Qwen2.5-Coder-0.5B & 54.7\% & 3.5\% & \textbf{0.0\%} & 52.1\% & -0.4\% & 0.8\% & 49.4\% & 0.6\% & 1.8\% \\
Qwen2.5-Coder-3B & 60.8\% & 1.3\% & 0.8\% & 53.1\% & 17.4\% & 1.2\% & 49.4\% & 20.7\% & \textbf{1.7\%} \\
Qwen2.5-Coder-7B & 63.5\% & 12.8\% & 0.3\% & 53.1\% & 53.5\% & 0.7\% & 52.6\% & 11.3\% & 19.1\% \\
Llama-3.2-1B & 39.2\% & 6.8\% & 0.8\% & 33.2\% & 6.8\% & 2.0\% & 35.3\% & 10.1\% & 3.9\% \\
Llama-3.2-3B & 44.2\% & 4.9\% & 1.7\% & 38.9\% & 35.5\% & 21.9\% & 32.7\% & 12.4\% & 19.5\% \\
\bottomrule
\end{tabular}
\caption{Category-wise performance for Error-prone, Performance, and Refactor alerts. Pass@1 is computed over samples having $\geq 1$ alert in the category; Reduction@P and New@P aggregate over passed instances.}
\label{app:category_performance_full}
\end{table*}

% Full Category-wise Performance (Style and Other)
\begin{table*}[t]
\centering
\small
\setlength{\tabcolsep}{6pt}
\begin{tabular}{lcccccc}
\toprule
 & \multicolumn{3}{c}{\textbf{Style}} & \multicolumn{3}{c}{\textbf{Other}} \\
\cmidrule(lr){2-4} \cmidrule(lr){5-7}
\textbf{Model} & Pass@1 & Reduction@P & New@P & Pass@1 & Reduction@P & New@P \\
\midrule
GPT-5.2 & 65.8\% & 91.4\% & \textbf{0.0\%} & 58.0\% & \textbf{100.0\%} & \textbf{0.0\%} \\
GPT-5.4-mini & \textbf{66.9\%} & \textbf{93.0\%} & \textbf{0.0\%} & 62.0\% & \textbf{100.0\%} & \textbf{0.0\%} \\
\midrule
DeepSeek-v3.2 & 55.5\% & 79.7\% & 1.0\% & 64.0\% & 98.4\% & \textbf{0.0\%} \\
Qwen2.5-Coder-32B & 61.3\% & 59.0\% & 0.5\% & 61.7\% & 95.3\% & \textbf{0.0\%} \\
Qwen2.5-Coder-0.5B & 46.5\% & -0.4\% & 2.2\% & 61.7\% & 3.0\% & \textbf{0.0\%} \\
Qwen2.5-Coder-3B & 50.3\% & 15.1\% & 1.4\% & \textbf{66.0\%} & 17.4\% & \textbf{0.0\%} \\
Qwen2.5-Coder-7B & 51.5\% & 28.1\% & 1.2\% & 61.7\% & 52.3\% & \textbf{0.0\%} \\
Llama-3.2-1B & 32.6\% & 9.2\% & 3.0\% & 40.4\% & 8.8\% & \textbf{0.0\%} \\
Llama-3.2-3B & 36.0\% & 19.7\% & 3.8\% & 55.3\% & 50.0\% & 2.3\% \\
\bottomrule
\end{tabular}
\caption{Category-wise performance (continued): Style and Other alerts under the same aggregation protocol as the first part.}
\label{app:category_performance_style_other}
\end{table*}

Tables~\ref{app:category_performance_full} and~\ref{app:category_performance_style_other} report performance across the five alert categories. Error-prone, Performance, and Refactor are grouped in the first table because they more often require behavior- or structure-aware changes; Style and Other appear in the second table. The split is presentational only: both tables use the same samples and aggregation protocol. Pass@1 is category-specific because it is computed on files containing at least one alert from that category, while Reduction@P and New@P aggregate category alerts over the corresponding passed revisions.

\paragraph{Error-prone alerts are hardest to fix while preserving functionality.} Error-prone alerts (e.g., \texttt{bugprone-*} in clang-tidy) require semantic understanding of the code's behavior. GPT-5.2 achieves 83.0\% Reduction@P on Error-prone with 63.3\% Pass@1, indicating that many fixes break functionality. Smaller models struggle more: \texttt{Qwen2.5-Coder-0.5B} achieves only 3.5\% Reduction@P on Error-prone, suggesting it either avoids fixing these alerts or fixes them incorrectly.

\paragraph{Performance alerts are easier to fix with localized edits.} Performance alerts (e.g., \texttt{performance-unnecessary-value-param}) often map to well-known idioms: pass-by-reference instead of pass-by-value, avoid unnecessary copies, use efficient data structures. GPT-5.4-mini achieves 90.3\% Reduction@P on Performance with 66.7\% Pass@1, one of the best balances among all categories. This suggests that performance fixes are more mechanical and less semantically risky.

\paragraph{Style alerts show cascading new-alert risks.} Style alerts are numerous but fine-grained. Fixing one Style violation (e.g., variable naming) can easily introduce another (e.g., inconsistent naming across the file). \texttt{Llama-3.2-3B} shows 3.8\% New@P on Style, among the highest for Style across models, confirming that Style fixes are prone to cascading violations.

\paragraph{Cross-category conclusion.} No category is uniformly easy across model scales. Commercial models achieve high Reduction@P in every category, but Refactor remains comparatively difficult for GPT-5.2 (59.3\%) and GPT-5.4-mini (72.5\%). For open-source models, Performance is often more tractable than Error-prone, while Style exposes a sharp scale gap despite its high frequency. The category results therefore support a two-part requirement for effective refactoring: semantic control for behavior-relevant alerts and consistent file-wide editing for numerous local Style alerts.

\subsection{Alert Count Analysis}
\label{sec:appendix_warning}

% ============================================================
% Placement: Appendix (Section A - Full Results)
% ============================================================
% A2: Full Alert Count Impact (1 and 2--5 alerts)
\begin{table*}[t]
\centering
\small
\setlength{\tabcolsep}{2pt}
\begin{tabular}{lcccccccc}
\toprule
& \multicolumn{4}{c}{\textbf{1 alert}} & \multicolumn{4}{c}{\textbf{2--5 alerts}} \\
\cmidrule(lr){2-5} \cmidrule(lr){6-9}
\textbf{Model} & Pass@1 & Reduction@P & New@P & LinePerFix@P & Pass@1 & Reduction@P & New@P & LinePerFix@P \\
\midrule
GPT-5.2 & 79.5\% & 64.5\% & 35.5\% & 21.9 & 71.7\% & 82.5\% & 10.0\% & 13.1 \\
GPT-5.4-mini & \textbf{82.1\%} & 62.5\% & 34.4\% & 12.9 & \textbf{75.1\%} & \textbf{83.0\%} & \textbf{8.1\%} & 9.1 \\
\midrule
DeepSeek-v3.2 & 61.5\% & \textbf{69.4\%} & 13.9\% & 17.1 & 63.0\% & 68.1\% & 9.5\% & 9.4 \\
Qwen2.5-Coder-32B & \textbf{82.1\%} & 46.9\% & 40.6\% & \textbf{11.2} & 70.5\% & 61.4\% & 16.6\% & 9.0 \\
Qwen2.5-Coder-0.5B & 59.0\% & 22.9\% & 11.4\% & 17.2 & 55.5\% & 8.7\% & 12.7\% & \textbf{8.9} \\
Qwen2.5-Coder-3B & 59.0\% & 48.6\% & \textbf{8.6\%} & 20.0 & 54.9\% & 33.0\% & 11.9\% & 9.9 \\
Qwen2.5-Coder-7B & 59.0\% & 62.9\% & \textbf{8.6\%} & 16.4 & 54.9\% & 51.1\% & 10.5\% & 9.6 \\
Llama-3.2-1B & 43.6\% & 25.9\% & 11.1\% & 20.8 & 39.3\% & 14.4\% & 11.5\% & 11.4 \\
Llama-3.2-3B & 56.4\% & 58.8\% & 14.7\% & 18.4 & 46.2\% & 30.5\% & 22.2\% & 12.4 \\
\bottomrule
\end{tabular}
\caption{Alert-count breakdown for files with 1 or 2--5 alerts. All four metrics are reported; best results per column are in \textbf{bold}.}
\label{app:alert_count_full}
\end{table*}

% Full Alert Count Impact (6--10 and >10 alerts)
\begin{table*}[t]
\centering
\small
\setlength{\tabcolsep}{2pt}
\begin{tabular}{lcccccccc}
\toprule
& \multicolumn{4}{c}{\textbf{6--10 alerts}} & \multicolumn{4}{c}{\textbf{$>$10 alerts}} \\
\cmidrule(lr){2-5} \cmidrule(lr){6-9}
\textbf{Model} & Pass@1 & Reduction@P & New@P & LinePerFix@P & Pass@1 & Reduction@P & New@P & LinePerFix@P \\
\midrule
GPT-5.2 & \textbf{60.0\%} & 87.2\% & 8.7\% & 13.7 & \textbf{64.5\%} & 89.0\% & 6.3\% & 19.3 \\
GPT-5.4-mini & \textbf{60.0\%} & \textbf{88.7\%} & 6.9\% & 10.1 & 63.9\% & \textbf{92.1\%} & \textbf{4.7\%} & 17.5 \\
\midrule
DeepSeek-v3.2 & 54.2\% & 75.5\% & 4.6\% & 12.7 & 53.6\% & 73.6\% & 6.4\% & 16.3 \\
Qwen2.5-Coder-32B & 55.0\% & 69.4\% & 8.3\% & 11.1 & 52.4\% & 53.1\% & 5.7\% & 16.1 \\
Qwen2.5-Coder-0.5B & 46.7\% & -2.8\% & 13.6\% & 11.1 & 41.0\% & -4.1\% & 8.9\% & \textbf{8.2} \\
Qwen2.5-Coder-3B & 51.7\% & 15.5\% & 10.6\% & 11.9 & 47.6\% & 0.1\% & 11.3\% & 14.7 \\
Qwen2.5-Coder-7B & 48.3\% & 27.4\% & 10.7\% & 13.1 & 53.0\% & 17.0\% & 13.8\% & 14.2 \\
Llama-3.2-1B & 25.8\% & 4.0\% & \textbf{4.0\%} & \textbf{4.7} & 31.3\% & -4.8\% & 17.0\% & 16.7 \\
Llama-3.2-3B & 32.5\% & 18.5\% & 18.8\% & 12.3 & 31.3\% & 15.2\% & 9.0\% & 12.6 \\
\bottomrule
\end{tabular}
\caption{Alert-count breakdown (continued): files with 6--10 or more than 10 alerts. The same metric definitions apply; best results per column are in \textbf{bold}.}
\label{app:alert_count_dense}
\end{table*}

% ============================================================
% Placement: Appendix (supports Section 4.4 mechanism analysis)
% ============================================================
\begin{table*}[t]
\centering
\small
\begin{tabular}{l rrrr rr r}
\toprule
& \multicolumn{4}{c}{\textbf{Fixed alerts by category}} & \textbf{\#Fix} & \textbf{\#New} & \textbf{Fix/New} \\
\cmidrule(lr){2-5}
\textbf{Format (3B)} & E-prone & Perf & Refac & Style & Total & Actual & Ratio \\
\midrule
Base & 8 & 47 & 39 & 254 & 360 & 261 & 1.4$\times$ \\
A+L$\rightarrow$A$'$ & 229 & 241 & 138 & 1803 & 2453 & 189 & 13.0$\times$ \\
Q$\rightarrow$A & 6 & 20 & 14 & 237 & 281 & 98 & 2.9$\times$ \\
Q$\rightarrow$A$'$ & 12 & 104 & 14 & 167 & 304 & 585 & 0.5$\times$ \\
\bottomrule
\end{tabular}
\caption{Per-category alert breakdown for 3B SFT models (Qwen2.5-Coder-3B). \emph{Fixed} counts denote original alerts removed per category. \#New is the total newly introduced alerts from the evaluation pipeline. The ``Other'' category is omitted ($\leq3.3\%$ of total fixed alerts). Fix/New Ratio shows how many alerts are fixed per new alert introduced.}
\label{app:sft_alert_category_breakdown}
\end{table*}

Tables~\ref{app:alert_count_full} and~\ref{app:alert_count_dense} present performance across four alert-count tiers. The first table covers sparse files with 1 or 2--5 alerts, and the second covers denser files with 6--10 or more than 10 alerts. Comparing the same model across the two tables reveals whether additional alerts provide useful opportunities for coordinated cleanup or instead overload the model's ability to preserve functionality and avoid collateral changes.

\paragraph{Alert density effects differ by model capability.} For stronger models, higher alert density reduces Pass@1 but increases Reduction@P: GPT-5.4-mini's Pass@1 drops from 82.1\% on single-alert files to 63.9\% on files with $>$10 alerts (a decline of 18.2 percentage points), while Reduction@P rises from 62.5\% to 92.1\%. Open-source models often degrade instead: \texttt{Qwen2.5-Coder-7B}'s Reduction@P drops from 62.9\% to 17.0\%, and \texttt{Llama-3.2-1B}'s from 25.9\% to $-$4.8\%. This suggests that stronger models adopt a broader editing strategy to resolve clustered alerts in bulk, while weaker models lack the capacity to coordinate multiple fixes.

\paragraph{Small models collapse on high-density files.} \texttt{Qwen2.5-Coder-0.5B} shows negative Reduction@P on files with $>$10 alerts (-4.1\%), meaning it introduces more alerts than it fixes. This indicates that small models lack the capacity to reason about multiple interacting violations and instead make scattered edits that create new problems.

\paragraph{New-alert introduction decreases with density for closed-source commercial models.} For GPT-5.2, New@P drops from 35.5\% on single-alert files to 6.3\% on files with $>$10 alerts; a similar trend holds for GPT-5.4-mini. This may reflect a selection effect: files with many alerts are likely already low-quality, so models have more room to improve without triggering new violations. This trend does not hold consistently for open-source models: \texttt{Qwen2.5-Coder-7B}'s New@P rises from 8.6\% to 13.8\%, and \texttt{Llama-3.2-1B}'s from 11.1\% to 17.0\%.

\paragraph{Conclusion.} Alert density separates coordinated refactoring from isolated fixes. Strong models trade some functionality preservation for substantially greater alert reduction as density rises, whereas several small models approach zero or negative Reduction@P in the densest bin. More alerts are therefore not merely more instances of the same task: they change the required behavior from a local repair to a coordinated file-level transformation, which current smaller models do not perform reliably.

\subsection{SFT Training Details}
\label{sec:appendix_sft_data}

% SFT Data Statistics
% Statistics of the training data for SFT experiments
\begin{table}[t]
\centering
\caption{Statistics of SFT training data. This table summarizes the revision-format training split only; it is kept in the appendix because the main text now focuses on the benchmark and evaluation tables.}
\label{app:sft_data_statistics}
\resizebox{\columnwidth}{!}{
\begin{tabular}{lcccc}
\toprule
\textbf{Language} & \textbf{\#Samples} & \textbf{Avg. Messages} & \textbf{Rule Types} & \textbf{Linter} \\
\midrule
C++ & 538 & 10.4 & 89 & clang-tidy \\
Python & 416 & 8.9 & 72 & Pylint \\
Java & 242 & 18.8 & 63 & PMD \\
\midrule
Total & 1,196 & 11.6 & 224 & -- \\
\bottomrule
\end{tabular}
}
\end{table}

Table~\ref{app:sft_data_statistics} presents the fine-tuning corpus: 1,196 problem instances sourced outside LintEval, with a distribution mirroring LintEval's language proportions (C++ 538, Python 416, Java 242). The average alert count per sample (11.6) is comparable to LintEval's average (13.1).

The corpus is separated from LintEval at the problem-instance level, so evaluation files are not reused as supervision examples. Matching the language proportions and approximate alert density reduces two obvious distribution shifts while leaving model family and supervision format as the controlled experimental factors. The corpus is not intended to reproduce every rule frequency exactly; instead, it supplies enough multilingual, multi-alert examples to compare how the same backbone responds to refactoring-oriented and generation-oriented objectives.

\paragraph{Per-Category Breakdown.}

Table~\ref{app:sft_alert_category_breakdown} decomposes fixed and newly introduced alerts for the \texttt{3B} SFT variants. \textbf{Note:} it reports \emph{raw alert counts} on the shared evaluation subset, \emph{not directly comparable} to the percentage metrics reported for the SFT experiment in the main paper. \textbf{A+L$\to$A$'$} fixes 2,453 alerts (vs.\ baseline 360) with 189 new alerts (Fix/New Ratio 13.0$\times$), the most efficient among all formats. \textbf{Q$\to$A} fixes only 281 alerts with 98 new alerts (Fix/New Ratio 2.9$\times$), indicating minimal lint reduction progress. \textbf{Q$\to$A$'$} fixes 304 alerts but introduces 585 new alerts, yielding the lowest Fix/New Ratio (0.5$\times$), suggesting that generation-oriented supervision with purified targets can produce unstable new-alert behavior on this subset.

The raw-count view complements, rather than replaces, Reduction@P and New@P. It exposes the volume of edits behind each aggregate percentage and makes the asymmetry between fixing and introducing alerts concrete. A+L$\to$A$'$ produces both the largest number of fixes and the strongest fix-to-new ratio, whereas Q$\to$A$'$ introduces more alerts than it removes on this subset. This supports the interpretation that access to the original code and its lint feedback is essential for learning targeted transformations; exposure to cleaner target programs alone does not provide the same localization signal.

\subsection{Qualitative Examples}
\label{sec:appendix_examples}

The examples in this section are not intended to estimate aggregate performance. Instead, they make the three failure and success modes measured by LintEval concrete: a useful revision should remove the reported alert, preserve the original behavior, and keep the edit localized enough to avoid collateral violations. Read together, the examples show why lint alert reduction alone is insufficient as a quality criterion. A revision can be syntactically plausible and even pass a limited test while still failing to implement the intended semantic change.

\paragraph{Example 1: Correct Minimal Revision (C++).}
\begin{tcolorbox}[colback=gray!5,colframe=gray!50,sharp corners]
\ttfamily\small
\textbf{Linter:} clang-tidy \texttt{performance-unnecessary-value-param} \quad
\textbf{Original:} \texttt{void process(vector<int> data) \{} \quad
\textbf{Revision:} \texttt{void process(const vector<int>\& data) \{} \quad
\textbf{Result:} Alert resolved, functionality preserved, single-line edit.
\end{tcolorbox}
\noindent A targeted edit (pass-by-value $\to$ pass-by-const-reference) that resolves the alert without side effects. This example supports the conclusion that many Performance alerts are actionable through a localized, idiomatic transformation: the model changes the parameter representation while leaving the function's observable behavior unchanged. It therefore illustrates the desired high-quality outcome measured jointly by positive Reduction@P, successful functionality tests, and low edit cost.

\paragraph{Example 2: Over-Editing (Python).}
\begin{tcolorbox}[colback=gray!5,colframe=gray!50,sharp corners]
\ttfamily\small
\textbf{Linter:} Pylint \texttt{unused-variable} \quad
\textbf{Original:} 15-line function with one unused variable. \quad
\textbf{Revision:} 42-line rewrite of the entire function. \quad
\textbf{Result:} Original alert resolved, tests pass, but three new Pylint alerts introduced.
\end{tcolorbox}
\noindent Instead of a targeted fix, the model rewrites the entire function, introducing new violations (e.g., naming conventions, missing docstrings). The conclusion is that over-editing creates a second source of risk even when the original alert disappears: a broad rewrite increases the surface area for new alerts and behavioral regressions. This is why New@P and LinePerFix@P complement Reduction@P, and why a model should not receive full credit for removing one alert if it does so by unnecessarily transforming the rest of the file.

\paragraph{Example 3: Superficial Fix (C++).}
\begin{tcolorbox}[colback=gray!5,colframe=gray!50,sharp corners]
\ttfamily\small
\textbf{Linter:} clang-tidy \texttt{bugprone-integer-division} \quad
\textbf{Original:} \texttt{int result = a / b;} (intended result is fractional). \quad
\textbf{Revision:} \texttt{float result = a / b;} \quad
\textbf{Result:} Integer division still performed; correct fix would promote an operand, e.g., \texttt{(float)a / b}.
\end{tcolorbox}
\noindent The model changes the storage type but misses that integer operands still produce integer division---illustrating why Error-prone alerts require semantic understanding beyond syntactic pattern matching. This example supports the conclusion that a superficially appropriate type change can leave the reported semantic defect untouched. Correctness depends on the types of the operands at the division operation, not only on the destination variable, so a reliable system must reason about evaluation semantics and test behavior rather than match the alert's wording or alter a nearby declaration.

\paragraph{Overall conclusion from the examples.}
Taken together, the three cases define a practical hierarchy of revision quality. The first is a minimal and semantically valid repair; the second removes an alert but incurs unnecessary collateral changes; the third appears plausible but fails to repair the underlying computation. They therefore motivate the benchmark's joint evaluation design: Reduction@P captures alert removal among functionally passing outputs, New@P captures collateral lint damage, and LinePerFix@P captures edit efficiency. The qualitative evidence also explains why Error-prone and mixed-category instances remain difficult even when a model can identify the reported rule. Successful linter-guided refactoring requires the model to select the right code location, apply the smallest behavior-preserving transformation, and verify that the resulting program still computes the intended result.

\end{document}